\documentclass[preprint,11pt]{elsarticle}
\usepackage[top=0.7in, right=1in, left=1in, bottom=1in]{geometry}
\usepackage{graphicx}%
\usepackage{multirow}%
\usepackage{amsmath,amssymb,amsfonts,amsthm, bm}%
\usepackage{empheq}%
\usepackage{mathrsfs}%
\usepackage[title]{appendix}%
\usepackage[dvipsnames]{xcolor}%
\usepackage{caption,subcaption,float}
\usepackage{placeins}
\usepackage{textcomp}%
\usepackage{hyperref,url}
\usepackage{manyfoot}%
\usepackage{booktabs}%
\usepackage{algorithm}%
\usepackage{pgfplots}
\usepackage{tikz}
\usepackage{natbib}
\usepackage{algorithmic}

\usepackage{hypernat}
\hypersetup{
    colorlinks=true,
    linkcolor=blue,
    citecolor=blue,
    filecolor=magenta,
    urlcolor=blue,
    pdfborder={0 0 0},
    pdfborderstyle={/S/U/W 0},
    pdftitle={DilectricPaper},
    pdfpagemode=FullScreen,
}

\AtBeginDocument{\color{black}}

\usepackage{xpatch}
\xpatchcmd{\citet}{\@citep}{}{}{}

\makeatletter
\def\ps@pprintTitle{%
 \let\@oddhead\@empty
 \let\@evenhead\@empty
 \def\@oddfoot{\reset@font\hfil\thepage\hfil}
 \let\@evenfoot\@oddfoot
}
\makeatother

\newtheoremstyle{remarkstyle}
  {3pt}
  {3pt}
  {\itshape}
  {}
  {\bfseries}
  {:}
  {.5em}
  {}

\theoremstyle{remarkstyle}

\newcommand{\rev}[1]{#1}

\def\bfn{{\bf n}}

\def\bfu{{\bf u}}

\def\bfx{{\bf x}}

\def\bfI{{\bf I}}

\def\bfM{{\bf M}}

\def\bfR{{\bf R}}

\def\bfX{{\bf X}}

\def\bfU{{\bf U}}

\def\bfw{\mbox{\boldmath $\omega$}}

\def\e0{\varepsilon_0}

\def\p{\partial}

\def\s0{\sigma_0}

\def\p{\partial}

\newcommand{\croch}[1]{{\left[ #1 \right]}}
\newcommand{\accol}[1]{{\left\{ #1 \right\}}}

\long\def\symbolfootnote[#1]#2{\begingroup%
\def\thefootnote{\fnsymbol{footnote}}\footnote[#1]{#2}\endgroup}

\pgfplotsset{compat=1.16}
\begin{document}

\begin{frontmatter}

\title{Wafer-to-Wafer Bonding: Part: I - The Coupled Physics Problem and the 2D Finite Element Implementation}

\author[1]{Kamalendu Ghosh\fnref{equal}}\ead{kamalendugho@micron.com, kamalendu.iitkgp@gmail.com}
\author[2]{Bhavesh Shrimali\fnref{equal}}\ead{bhavesh.shrimali@gmail.com}
\author[3]{Subin Jeong}\ead{sooibing@tamu.edu}

\fntext[equal]{Kamalendu Ghosh and Bhavesh Shrimali contributed equally to this work.}
\address[1]{Visiting Researcher, Aerospace Engineering, University of Illinois at Urbana-Champaign, Illinois 61801, USA}
\address[2]{University of Illinois at Urbana-Champaign, Illinois 61801, USA}
\address[3]{Department of Mechanical Engineering, Texas A \& M University, College Station, TX 77840}

\begin{abstract}
Wafer-to-wafer (WxW) bonding is a key enabler for three-dimensional integration, including hybrid bonding for fine-pitch Cu--Cu interconnects. During bonding, wafer deformation and the air entrapped between the wafers interact through a strongly coupled, time-dependent fluid--structure interaction (FSI) that can produce non-intuitive bonding dynamics and process sensitivities. This paper develops a mathematically consistent reduced-order model for WxW bonding by deriving a Kirchhoff--Love plate equation for wafer bending from three-dimensional linear elasticity and coupling it to a Reynolds lubrication equation for the inter-wafer air film. The resulting nonlinear plate--Reynolds system is discretized and solved monolithically in the high-performance FEniCSx framework using a $C^0$ interior-penalty formulation for the fourth-order plate operator, standard continuous Galerkin discretization for the pressure field, implicit time integration, and a Newton solver with automatic differentiation. Simulations reproduce experimentally reported wafer displacement histories for multiple initial gaps and verify force equilibrium at the bond front, where the Reynolds pressure acts as an effective contact reaction. Parametric studies reveal nonlinear, and in some cases non-monotonic, sensitivities of bonding-front kinetics to the initial gap, air viscosity, and interfacial energy, providing actionable trends for process optimization.
\end{abstract}


\end{frontmatter}

\section{Introduction}\label{sec:introduction}

Wafer-to-Wafer (WxW) bonding has emerged as a cornerstone technology for next-generation semiconductor packaging, enabling three-dimensional integrated circuit (3D-IC) stacking, High Bandwidth Memory (HBM), MEMS fabrication, and photonic device integration~\cite{bao2021review, moriceau2010physics}, \rev{\cite{gosele1998semiconductor}}. Unlike traditional bump-based stacking, which introduces inter-die gaps of approximately $30\,\mu$m and is limited to interconnect pitches above $10\,\mu$m, (hybrid) WxW bonding achieves direct Cu-to-Cu connections with pitches below $10\,\mu$m, enabling substantially higher I/O density, reduced package dimensions, lower electrical resistance and superior thermal performance~\cite{bao2021review}, \rev{\cite{wang2023hybrid}}. These attributes translate directly into higher bandwidth, improved signal integrity, reduced power consumption, and enhanced heat dissipation, all critical requirements for enabling miniaturization of packages and modern HBM architectures. Despite this transformative potential, WxW bonding involves high process sensitivities, such as - submicrometer particulate contamination can nucleate voids at the bond interface, leading to delamination, incomplete bonding, high overlays, and degraded mechanical and electrical integrity. Furthermore, the governing physics of the bonding process involve a strongly coupled, nonlinear fluid--structure interaction that can give rise to non-intuitive behaviors, such as non-monotonic relationships between overlay and initial bond gap. Developing a rigorous predictive model of the bonding process is therefore not merely of academic interest but of direct industrial relevance, as it provides a pathway to understanding and mitigating these failure mechanisms.

\rev{The mechanics of wafer bonding were first analyzed using finite element methods by Turner and Spearing~\cite{turner2002modeling}, who established the plate-bending framework for predicting bonding-induced deformations.} The dynamics of the bonding front have been studied since the seminal work of Rieutord~et~al.~\cite{rieutord2005dynamics}, who proposed a model relating the bonding front velocity to the surface adhesion energy and the viscous drag of air escaping the shrinking gap between the wafers. \rev{Experimental measurements of adhesion energies and bonding wave velocities were subsequently reported by Larrey~et~al.~\cite{larrey2016adhesion}.} This competition between elastic driving forces and viscous resistance was later formalized as a coupled Fluid--Structure Interaction (FSI) problem by Navarro~et~al.~\cite{navarro2013direct}, wherein the elastic deformation of the wafer governs the air pressure between the wafers and the resulting air pressure field in turn loads the wafer. The dynamics of the bonding wave has also been analytically examined by Radisson~et~al.~\cite{radisson2015modelling}, who modeled the propagation speed of the bonding front as a function of the material and geometric parameters. \rev{More recently, Jain~\cite{jain2026bond} derived a self-consistent analytical formulation for the bond front velocity by re-parameterizing the viscous dissipation region, providing power-law solutions for the substrate shape during bonding.} More recent efforts have employed finite element methods to model various aspects of the bonding process, including bonding wave propagation and cavity-induced deformation modes~\cite{ip2022multi, lim2020design, tang2025numerical}, thermo-mechanical effects in glass-frit bonding~\cite{farshchi2023experimental}, and Cu pad expansion in hybrid bonding~\cite{tsau2022simulation}. Together, these studies have demonstrated the viability of the FSI framework for capturing the essential physics of WxW bonding.

Despite this body of work, existing models share three common limitations: \rev{(a) the plate equations employed are not derived from three-dimensional elasticity with explicit statement of the Kirchhoff--Love kinematic hypotheses and their validity criteria; (b) the complete weak formulations necessary for finite element implementation---including the treatment of the fourth-order plate operator, boundary conditions, and discretization details---are not provided, limiting reproducibility; and (c) the inherently nonlinear nature of the coupled system, which gives rise to non-intuitive phenomena such as non-monotonic dependence of bonding wave speed on initial gap, has not been systematically analyzed or attributed to the mathematical structure of the governing equations.} Previous formulations generally rely on simplified plate models that are not rigorously derived from the full three-dimensional governing equations of linear elasticity, and they often do not clearly articulate the assumptions under which such dimensional reductions are justified~\cite{navarro2013direct, ip2022multi}. This gap is significant because the wafer–wafer bonding problem is inherently three-dimensional, involving aspects such as asymmetric incoming wafer geometries, the anisotropic behavior of silicon, and asymmetric boundary conditions driven by how the wafers are constrained and subsequently released. In most cases, the three-dimensional balance of linear momentum, the Kirchhoff–Love kinematic assumption, and the derivation of the flexural rigidity are neither made explicit nor tied to a properly posed boundary value problem. Furthermore, the \emph{nonlinear nature} of the coupled system has been largely overlooked: the two-way coupling
between wafer deflection and air pressure (between the two wafers) yields a genuinely nonlinear, time-dependent FSI problem whose solution behavior is far from intuitive. Non-intuitive phenomena---such as pressure-induced retardation of the bond front and the non-monotone dependence of bonding wave speed on initial bond gap---have not been rigorously analyzed or attributed to the structure of the governing equations. 

The present work addresses these shortcomings by developing a mathematically rigorous, fully coupled FSI formulation for WxW bonding, implemented in the high-performance FEniCSx finite element framework~\cite{baratta2023dolfinx, ghosh2025hybrid}. Starting from the three-dimensional equations of linear elasticity for an isotropic wafer, we perform a systematic reduction to the classical Kirchhoff--Love plate equation via through-thickness integration~\cite{shrimali2021remarkable}, making all assumptions and their validity criteria explicit. The air between the wafers is governed by the Reynolds lubrication equation, and the two-way coupling -- wherein the wafer deflection and its rate determine the air pressure, which in turn acts as a distributed transverse load on the wafer -- is preserved in full. \rev{Notably, the Reynolds pressure field in the bonded region naturally serves as a contact reaction that resists interpenetration of the wafers, eliminating the need for a separate contact algorithm.} The resulting \emph{inherently nonlinear} coupled equations, and the results presented in Section~\ref{sec:results} demonstrate several
counter-intuitive implications of this nonlinearity that have not been previously reported or discussed in the WxW bonding literature. The implementation leverages the \textsc{FEniCSx} framework~\cite{baratta2023dolfinx, scroggs2022construction, scroggs2022basix, UFL}, which has proven to be effective for analogous nonlinear multiphysics problems, including nonlinear viscoelasticity~\cite{ghosh2021nonlinear, shrimali2023nonlinear} and dielectric elastomers~\cite{ghosh2021two}.

\rev{This paper is the first in a planned series. Part~I is deliberately restricted to the reduced two-dimensional problem, which suffices for the bonding-front kinetics and air-film pressure studied here, but excludes effects that are intrinsically three-dimensional: in-plane deformation and the resulting post-bond overlay, anisotropic flexural response, asymmetric incoming wafer geometries, and lift-off of the bottom wafer from its chuck, which the present rigid, perfectly adhered bottom wafer precludes by construction. Later parts will progressively address these restrictions, with the eventual objective of solving the full three-dimensional bonding problem. Since the coupling structure established here is unchanged by such extensions, the formulation and monolithic solution strategy of Part~I carry over directly; see Section~\ref{sec:future}.}

The remainder of this paper is organized as follows.
Section~\ref{sec:theory} presents the complete mathematical formulation, beginning with the three-dimensional governing equations and proceeding through their systematic
reduction to the two-dimensional coupled plate -- Reynolds system. Section~\ref{sec:fenicsx} describes finite element discretization and FEniCSx implementation. Section~\ref{sec:results} presents numerical results, including validation studies and the non-intuitive physical phenomena arising from the nonlinear coupling. Section~\ref{sec:future} concludes with a summary and directions for future work.

\section{Theory}\label{sec:theory}
The mechanics governing Wafer-to-Wafer (WxW) bonding is characterized by a coupled Fluid-Structure Interaction (FSI) problem. Consider, two wafers in their undeformed initial configurations at time, $t=0$, within a Cartesian coordinate system, positioned coaxially by the bonding apparatus on the top and bottom chuck occupying domains $\Omega_{top}$ and $\Omega_{bot}$ respectively, such that $\Omega_{top} = \{\bfX \in \mathbb{R}^3: X_1^2 + X_2^2 \leq R^2, -t_w \leq X_3 \leq 0\}$ and $\Omega_{bot} = \{\bfX \in \mathbb{R}^3: X_1^2 + X_2^2 \leq R^2, -2t_w-h_0 \leq X_3 \leq -t_w - h_0\}$. The top wafer's center is located at (0,0,-$t_w/2$), and the center of the top surface of the bottom wafer is at (0,0,-$3t_w/2 -h_{0}$), resulting in an initial separation distance $h_0$ between the bottom surface of the top wafer and the top surface of the bottom wafer. A thin layer of air is entrapped within the interstitial gap between the wafers. At this stage, it is useful to define the following regions, as illustrated in Fig.~\ref{fig:wafer_setup}

\begin{align}
    \begin{cases}
    \partial \Omega^u_{top} = \{\bfX : X_1^2 + X_2^2 \leq R^2, X_3 = 0\}, & \text{Top surface of the Upper wafer} \\
    \partial \Omega^l_{top} = \{\bfX : X_1^2 + X_2^2 \leq R^2, X_3 = -t_w\}, & \text{Bottom surface of the Upper wafer} \\
    \partial \Omega^{lat}_{top} = \{\bfX : X_1^2 + X_2^2 = R^2, 0 \leq X_3 \leq -t_w\}, & \text{Lateral curved surface of the Upper wafer} \\
    \partial \Omega^u_{bot} = \{\bfX : X_1^2 + X_2^2 \leq R^2, X_3 = -t_w - h_0 \}, & \text{Top surface of bottom wafer} \\
    \partial \Omega^l_{bot} = \{\bfX : X_1^2 + X_2^2 \leq R^2, X_3 = -2t_w - h_0 \},  & \text{Bottom surface of the wafer on the bottom.}
\end{cases}
\end{align}

\begin{figure}[htbp]
\centering
\begin{subfigure}[t]{\textwidth}
\centering
\begin{tikzpicture}[>=stealth, font=\small]
  \fill[blue!10] (-4,0) rectangle (4,-0.8);
  \draw[thick] (-4,0) rectangle (4,-0.8);
  \node at (-2.5,-0.4) {\large $\Omega_{top}$};

  \fill[orange!6] (-4,-0.8) rectangle (4,-1.4);
  \draw[densely dashed, orange!40!black] (-4,-0.8) -- (4,-0.8);
  \draw[densely dashed, orange!40!black] (-4,-1.4) -- (4,-1.4);
  \node[orange!50!black, font=\footnotesize] at (-2.5,-1.1) {Air ($p_{air}$)};

  \fill[teal!8] (-4,-1.4) rectangle (4,-2.2);
  \draw[thick] (-4,-1.4) rectangle (4,-2.2);
  \node at (-2.5,-1.8) {\large $\Omega_{bot}$};

  \draw[dotted, gray] (4,0) -- (5.8,0);
  \node[right, font=\scriptsize] at (5.8,0) {$\partial\Omega^u_{top}$~($X_3\!=\!0$)};
  \draw[dotted, gray] (4,-0.8) -- (5.8,-0.8);
  \node[right, font=\scriptsize] at (5.8,-0.8) {$\partial\Omega^l_{top}$~($X_3\!=\!{-}t_w$)};
  \draw[dotted, gray] (4,-1.4) -- (5.8,-1.4);
  \node[right, font=\scriptsize] at (5.8,-1.4) {$\partial\Omega^u_{bot}$};
  \draw[dotted, gray] (4,-2.2) -- (5.8,-2.2);
  \node[right, font=\scriptsize] at (5.8,-2.2) {$\partial\Omega^l_{bot}$~($X_3\!=\!{-}2t_w{-}h_0$)};

  \draw[|<->|] (-4.7,-0.8) -- node[left, font=\footnotesize]{$h_0$} (-4.7,-1.4);
  \draw[|<->|, gray] (-4.7,0) -- node[left, font=\footnotesize, gray]{$t_w$} (-4.7,-0.8);
  \draw[|<->|, gray] (-4.7,-1.4) -- node[left, font=\footnotesize, gray]{$t_w$} (-4.7,-2.2);
  \draw[<->, gray] (-4,-2.55) -- node[below, font=\footnotesize]{$2R$} (4,-2.55);

  \fill[red!15] (-0.6,0) rectangle (0.6,0.08);
  \draw[red!60!black, thick] (-0.6,0.08) -- (0.6,0.08);
  \draw[->, red!60!black, thick] (0,0.55) -- (0,0.12);
  \node[red!60!black, font=\footnotesize] at (0,0.9) {$f_S$};
  \draw[<->, red!50!black, thin] (-0.6,-0.12) -- node[below, font=\scriptsize]{$R_S$} (0.6,-0.12);

  \node[blue!50!black, font=\scriptsize, rotate=90] at (-4.22,-0.4) {$\partial\Omega^{lat}_{top}$};

  \draw[->, thick] (-5.3,0.55) -- (-4.5,0.55) node[right, font=\footnotesize]{$X_1$};
  \draw[->, thick] (-5.3,0.55) -- (-5.3,1.5) node[above, font=\footnotesize]{$X_3$};
\end{tikzpicture}
\caption{Cross-sectional view ($X_1$--$X_3$ plane) of the initial undeformed configuration.
  The top wafer $\Omega_{top}$ and bottom wafer $\Omega_{bot}$ are separated by an air gap of height~$h_0$.
  Boundary surfaces $\partial\Omega^{u,l}_{top}$ and $\partial\Omega^{u,l}_{bot}$ are labeled.}
\label{fig:domain_cross_section}
\end{subfigure}

\vskip 8pt

\begin{subfigure}[t]{\textwidth}
\centering
\input{Figures/fig_bonding_stages}
\caption{Schematic of the wafer-to-wafer bonding process (cross-sectional view):
  (a)~initial configuration with air gap $h_0$ between wafers $\Omega_{top}$ and $\Omega_{bot}$;
  (b)~a striker applies force $f_{\texttt{S}}$ to initiate contact at the center;
  (c)~bond front begins to propagate radially outward from the contact zone;
  (d)~bond front continues to advance while entrapped air is expelled;
  (e)~the top wafer is released from the chuck and the striker is retracted;
  (f)~full bonding across the entire interface.}
\label{fig:bonding_stages}
\end{subfigure}
\caption{(a)~Cross-sectional view of the initial domain configuration showing the top wafer $\Omega_{top}$
and the bottom wafer $\Omega_{bot}$ separated by an air gap of height~$h_0$, and
(b)~schematic of the six stages of the wafer-to-wafer bonding process.}
\label{fig:wafer_setup}
\end{figure}

Clearly, $\Omega_{top} \subset \mathbb{R}^3$ denotes the three-dimensional reference domain occupied by the top wafer. For the sake of completeness for discussions in the following sections, we introduce
\begin{align}
\omega = \{\bar{\bfx} = (X_1,X_2) \in \mathbb{R}^2 : X_1^2 + X_2^2 \leq R^2\}
\label{eq:omega_def}
\end{align}
as the two-dimensional mid-surface of the top wafer, obtained by projecting $\Omega_{top}$ onto the plane $X_3 = 0$.\footnote{The three-dimensional governing equations are posed on $\Omega_{top}$, while the reduced plate model operates on $\omega$ as will be described in Section~\ref{subsec:kl_reduction}.}

The subsequent bonding process encompasses the following four stages:

\begin{itemize}
    \item Initiation: The process commences when a striker of radius $R_{S}$ applies a localized force, $\mathbf{f}_\texttt{S}(\bfX,t) = f_{\texttt{S}}(\bfX,t)\mathbf{e}_3$ at time $t = 0$, where $\bfX \in \partial \Omega_{S} = \{\bfX \in \partial \Omega^{u}_{top}: X_1^2 + X_2^2 \leq R_{S}^2, X_3  = 0\}$, to the central region of the top wafer. This applied force induces a deformation in the top wafer, causing it to make initial physical contact with the bottom wafer at the point of maximum deflection and thereby modifying the effective separation $h(\bfX,t)$ between the wafers from its initial value $h(\bfX,t = 0) = h_0$.

    \item Bond Propagation: Following initiation, the bond front propagates radially outward. The interfacing surfaces of both wafers undergo a pretreatment process (e.g., plasma activation) to establish attractive interfacial forces $f_{\texttt{B}}(\bfX, h(\bfX,t))$. These forces become significant when the local separation between the wafer surfaces reduces below a critical threshold $\delta$ such that
    
    \begin{align}
        f_{\texttt{B}}(\bfX,h(\bfX,t)) = f_{\texttt{B}}(\bfX,h)=\begin{cases}
        = f_{\texttt{BE}}, & \text{if } h(\bfX,t) \leq \delta \\
        = 0, & \text{otherwise}.
        \end{cases}\label{eq:BE}
    \end{align}
    
    Consequently, these attractive forces draw the top wafer progressively towards the bottom wafer, concurrently expelling the entrapped air from the advancing bond interface. This \textit{constitutive assumption} is adopted throughout the remainder of the paper.

    \item Wafer Release: Subsequently, the top wafer is released from the retaining chuck of the bonding apparatus. This release allows the remainder of the top wafer to settle onto the bottom wafer under the influence of the propagating interfacial forces and potentially gravitational effects, facilitating the continued expansion of the bonded area.

    \item Bond Completion: The process culminates when the bond interface has propagated across the entire surface of the wafers, resulting in a fully bonded wafer pair.
\end{itemize}

\subsection{The three-dimensional governing equations}\label{subsec:3d_bvp}

Under the assumption of a rigid bottom wafer, the deformation of the top wafer is described by the displacement field
\begin{align}
\bfu(\bfX,t) = \big(u_1(\bfX,t),\, u_2(\bfX,t),\, w(\bfX,t)\big), \qquad \bfX \in \Omega_{top},\; t \in (0,T],
\label{eq:u_3d}
\end{align}
where $u_1$, $u_2$ denote the in-plane components and $w$ is the out-of-plane (transverse) component. The linearized strain tensor associated with \eqref{eq:u_3d} is
\begin{align}
\bm{\varepsilon}(\bfu) = \frac{1}{2}\!\left(\nabla \bfu + \nabla \bfu^\top\right).
\label{eq:strain}
\end{align}
For an isotropic, linearly elastic material, the Cauchy stress tensor is given by the constitutive relation
\begin{align}
\bm{\sigma} = \bm{\sigma}\!\left(\bm{\varepsilon}(\bfu)\right) = 2G\,\bm{\varepsilon}(\bfu) + \lambda\,\mathrm{tr}\!\left(\bm{\varepsilon}(\bfu)\right)\bfI,
\label{eq:constitutive}
\end{align}
where $G = E/[2(1+\nu)]$ and $\lambda = E\nu/[(1+\nu)(1-2\nu)]$ are the Lam\'{e} parameters, $E$ is Young's modulus, and $\nu$ is Poisson's ratio.

Under quasi-static conditions, the balance of linear momentum for the top wafer, incorporating all body and surface forces, reads
\begin{subequations}\label{eq:blm}
\begin{align}
\textrm{div}\,\bm{\sigma}(\bfX,t) &= \mathbf{b}(\bfX,t)
\quad \forall\, (\bfX,t) \in \Omega_{top}\times(0,T],
\label{eq:blm_a}\\
\bm{\sigma}\,\bfn &=
\begin{cases}
\mathbf{0}, & \forall\,(\bfX,t) \in \p\Omega^{lat}_{top}\times(0,T], \\
f_{\texttt{S}}(\bfX,t)\mathbf{e}_3, & \forall\,(\bfX,t) \in \p\Omega_{S}\times(0,T], \\
\mathbf{f}^{\texttt{C}}(\bfX,t), & \forall\,(\bfX,t) \in \p\Omega^{u}_{top}\times(0,T], \\
\mathbf{f}_{\texttt{C}}(\bfX,t) + \mathbf{f}_{\texttt{B}}(\bfX,t) + p(\bfX,t)\mathbf{n}(\bfx,t), & \forall\,(\bfX,t) \in \p\Omega^{l}_{top}\times(0,T],
\end{cases}
\label{eq:bc_lat}
\end{align}
\end{subequations}
where:
\begin{itemize}
\item $\mathbf{b}(\bfX,t)$ is the body-force density (e.g., $\mathbf{b} = \{0,0,b_3\} = -\rho g\,\mathbf{e}_3$),
\item $\mathbf{f}^{\texttt{C}}(\p\Omega^u_{top},t)$ is the contact traction exerted by the chuck on the top surface of the bonding apparatus,
\item $\mathbf{f}_{\texttt{B}}(\bfx,h)$ is the bonding traction (cf.\ \eqref{eq:BE}),
\item $\mathbf{f}_{\texttt{C}}(\bfX,t)$ is the contact traction preventing inter-penetration of the wafers,
\item $p(\bfX,t)$ is the pressure of the air in the bond gap, which induces the normal load $p(\bfX,t)\,\mathbf{n}(\bfX,t)$ on $\p\Omega^l_{top}$.
\end{itemize}

The air pressure $p$ is determined by the three-dimensional Reynolds lubrication equation posed on the gap domain $\Omega_{gap}$ with gap height $h(\bfX,t)=h_0+w(\bfX,t)$:
\begin{align}
12\mu\,\frac{\p h}{\p t} - \nabla\cdot\!\left(h^3\,\nabla p\right) = 0,
\qquad \forall\, \bfX \in \Omega_{gap},\; t \in (0,T],
\label{eq:reynolds_3d}
\end{align}
where $\mu$ is the dynamic viscosity of air.

\rev{The use of the Reynolds lubrication equation is justified by the following conditions, all of which are satisfied in the wafer bonding configuration~\cite{hamrock2004fundamentals}:
\begin{enumerate}
    \item[(i)] \textit{Thin-film assumption:} The gap height is much smaller than the lateral extent, $\varepsilon = h_0/R \sim 100\,\mu\text{m}/150\,\text{mm} \approx 6.7 \times 10^{-4} \ll 1$ (assuming a large gap of $100\,\mu\text{m}$).
    \item[(ii)] \textit{Low Reynolds number:} Viscous forces dominate over inertial forces. Taking the bond-front velocity $V \sim 10$~mm/s as the characteristic velocity, $\text{Re} = \rho_{\text{air}} V h_0 / \mu \approx (1.2)(0.01)(10^{-4})/(1.81 \times 10^{-5}) \approx 0.07 \ll 1$.
    \item[(iii)] \textit{Quasi-steady pressure:} The acoustic time scale for pressure equilibration, $\tau_{\text{acoustic}} \sim h_0/c_{\text{sound}} \approx 0.3\,\mu$s, is orders of magnitude smaller than the mechanical time scale ($\sim 1$--$10$~s).
    \item[(iv)] \textit{Continuum validity:} The Knudsen number $\text{Kn} = \lambda/h$, where $\lambda$ denotes the molecular mean free path of the gas, classifies the flow regime: $\text{Kn} < 0.01$ corresponds to continuum flow, $0.01 < \text{Kn} < 0.1$ to the slip-flow regime, and $\text{Kn} > 0.1$ to the transitional regime where the continuum-based Reynolds equation breaks down~\cite{burgdorfer1959influence, fukui1988analysis}. For air at atmospheric pressure and room temperature, $\lambda \approx 68\,\text{nm}$. At the initial gap $h_0 = 100\,\mu$m, $\text{Kn} \approx 6.8 \times 10^{-4} \ll 0.01$, placing the flow well within the continuum regime. As the gap closes during bonding, the local Knudsen number increases; however, the simultaneously rising air pressure near the bond front reduces $\lambda$ (since $\lambda \propto 1/P$ from kinetic theory)~\cite{navarro2014thesis}, partially mitigating the increase in $\text{Kn}$. The regularization introduced in Section~\ref{sec:fenicsx} ensures that the effective gap $\hat{h}$ remains sufficiently large to prevent the flow from entering the transitional regime.
\end{enumerate}}

The coupled three-dimensional FSI problem can thus be stated abstractly as:
\begin{subequations}\label{eq:3d_system}
\begin{align}
\mathcal{G}_1(\bfu,\,p) &\;\equiv\; \textrm{div}\,\bm{\sigma}(\bfu)
- \mathbf{b}
= \mathbf{0},
\label{eq:G1} \\[6pt]
\mathcal{G}_2(p,\,\bfu) &\;\equiv\; 12\mu\,\frac{\p h}{\p t} - \nabla\cdot\!\left(h^3\,\nabla p\right) = 0,
\label{eq:G2}
\end{align}
\end{subequations}
subjected to the traction boundary conditions in Eq.~\eqref{eq:bc_lat}. Here, the coupling is two-way: the displacement $\bfu$ determines the gap $h$ appearing in $\mathcal{G}_2$, and the pressure $p$ enters as a surface traction in $\mathcal{G}_1$. This system of governing equations shares structural similarities with models for the nonlinear viscoelasticity of rubber-like solids \cite{ghosh2021nonlinear, shrimali2023nonlinear} and the transient response of dielectric elastomers \cite{ghosh2021two}. Such problems have been successfully implemented using the legacy FEniCS computing platform and, more recently, the high-performance FEniCSx framework \cite{ghosh2025hybrid}.
\begin{table}[htbp]
\centering
\caption{Parameters appearing in Section~\ref{sec:theory}.}
\label{tab:theory_params}
\renewcommand{\arraystretch}{1.15}
\begin{tabular}{p{0.22\linewidth}p{0.70\linewidth}}
\hline
\textbf{Symbol} & \textbf{Meaning}\\
\hline
\multicolumn{2}{l}{\textbf{Geometric/kinematic parameters}}\\
$\bfX$ & Reference (material) coordinate in $\mathbb{R}^3$; $t$ denotes time.\\
$\bar{\bfx}$ & In-plane coordinate on the mid-surface $\omega$.\\
$\Omega_{top}$ & Three-dimensional reference domain of the top wafer.\\
$\Omega_{bot}$ & Three-dimensional reference domain of the bottom wafer.\\
$\Omega_{gap}$ & Thin interstitial air-gap domain between the wafers.\\
$\omega$ & Two-dimensional mid-surface domain of the top wafer (projection of $\Omega_{top}$ onto $X_3=0$).\\
$\partial\Omega^{u,l}_{top}$ & Top/bottom surfaces of the top wafer; $\partial\Omega^{lat}_{top}$ its lateral surface.\\
$\Gamma_R$ & Outer edge of the plate/mid-surface, $\Gamma_R=\{\bar{\bfx}\in\mathbb{R}^2: X_1^2+X_2^2=R^2\}$.\\
$\Gamma_{\texttt{vent}}$ & Vented portion of the boundary where the air gap is open to ambient; here $\Gamma_{\texttt{vent}}=\Gamma_R$.\\
$\mathbf{n}$ & Outward unit normal.\\
$R$ & Wafer radius.\\
$t_w$ & Wafer thickness.\\
$h_0$ & Initial air-gap height (initial wafer separation).\\
$h(\bfX,t)$ & Local gap between wafers (here $h=h_0+w$).\\
$r_b(t)$ & Bond-front radius, i.e. the radial position of the moving bonded/unbonded interface.\\
$R_{\texttt{S}}/R_{pin}$ & Striker/pin radius.\\
$T$ & Final time of interest for $t\in(0,T]$.\\
$\bfu(\bfX,t)$ & Displacement field; $w$ denotes the transverse displacement component.\\
$\bm{\varepsilon}(\bfu)$ & Linearized strain tensor.\\
$\bm{\sigma}$ & Cauchy stress tensor.\\[2pt]
\multicolumn{2}{l}{\textbf{Constitutive parameters}}\\
$E,\,\nu$ & Young's modulus and Poisson's ratio.\\
$G,\,\lambda$ & Lam\'{e} parameters.\\
$D$ & Plate flexural rigidity, $D=Et_w^3/[12(1-\nu^2)]$.\\
$\rho$ & Mass density.\\
$\mu$ & Dynamic viscosity of air.\\[2pt]
\multicolumn{2}{l}{\textbf{Force/loading parameters}}\\
$\mathbf{b}(\bfX,t)$ & Body-force density (e.g., gravity, $\mathbf{b}=-\rho g\,\mathbf{e}_3$).\\
$g$ & Gravitational acceleration.\\
$\mathbf{f}^{\texttt{C}}(\bfX,t)$ & Chuck contact traction applied on $\partial\Omega^{u}_{top}$.\\
$\mathbf{f}_{\texttt{C}}(\bfX,t)$ & Unilateral contact traction preventing inter-penetration.\\
$\mathbf{f}_{\texttt{S}}(\bfX,t)=f_{\texttt{S}}(\bfX,t)\mathbf{e}_3$ & Striker-applied surface traction/force used to initiate contact.\\
$q_{\texttt{ext}}(\bar{\bfx},t)$ & Resultant transverse external loading per unit area of the mid-surface.\\
$\delta$ & Critical separation below which bonding forces activate.\\
$\mathbf{f}_{\texttt{B}}(\bfX,h)$ & Interfacial bonding traction as a function of gap.\\
$f_{\texttt{BE}}$ & Constant bonding traction for $h\le \delta$.\\
$\Gamma = f_{\texttt{BE}}\delta$ & Constant bonding energy $h\le \delta$.\\
$p(\bfX,t)$ & Pressure of the entrapped air in the gap.\\
$p_{\texttt{atm}}$ & Ambient (atmospheric) pressure prescribed on $\Gamma_{\texttt{vent}}$.\\
\hline
\end{tabular}
\end{table}

\subsection{Reduction to the Kirchhoff--Love plate model}\label{subsec:kl_reduction}

Following \cite{shrimali2021remarkable}, the three-dimensional problem $\mathcal{G}_1 = \mathbf{0}$, $\mathcal{G}_2 = 0$ and the traction boundary conditions in Eq.~\eqref{eq:bc_lat} can be reduced to a two-dimensional plate model in the thin-wafer limit $t_w/R \ll 1$. The classical Kirchhoff--Love hypothesis asserts:
\begin{enumerate}
\item[(\textit{i})] Straight material fibers initially normal to the mid-surface remain straight and normal to the deformed mid-surface (i.e., the transverse shear strains vanish, $\varepsilon_{\alpha 3} = 0$, $\alpha = 1,2$).
\item[(\textit{ii})] The thickness of the plate does not change during deformation (i.e., $\varepsilon_{33} = 0$).
\end{enumerate}

\noindent\textit{Kinematics.}\; Under these hypotheses, the three-dimensional displacement field \eqref{eq:u_3d} reduces to the form
\begin{align}
u_\alpha(\bfX,t) = -X_3\,\frac{\p w}{\p X_\alpha}(\bar{\bfx},t), \quad \alpha = 1,2, \qquad u_3(\bfX,t) = w(\bar{\bfx},t),
\label{eq:kl_disp}
\end{align}
where $\bar{\bfx} = (X_1, X_2)$ are the in-plane coordinates of the mid-surface $\omega$ (cf.\ \eqref{eq:omega_def}) and $w(\bar{\bfx},t)$ is the transverse deflection of the mid-surface. In particular, the in-plane displacements are entirely determined by $w$ and its gradient, so that at the mid-surface ($X_3 = 0$)
\begin{align}
\bfu \longrightarrow \big(0,\,0,\,w(\bar{\bfx},t)\big).
\label{eq:plate_limit_u}
\end{align}
Substituting \eqref{eq:kl_disp} into the strain definition \eqref{eq:strain}, the only non-vanishing components of $\bm{\varepsilon}$ are
\begin{align}
\varepsilon_{\alpha\beta} = -X_3\,\kappa_{\alpha\beta}(\bar{\bfx},t), \qquad
\kappa_{\alpha\beta} = \frac{\p^2 w}{\p X_\alpha\,\p X_\beta}, \quad \alpha,\beta = 1,2,
\label{eq:kl_strain}
\end{align}
where $\bm{\kappa} = \nabla^2 w$ is the curvature tensor of the mid-surface.

\noindent\textit{Bending moments.}\; Invoking the plane-stress condition ($\sigma_{33} = 0$) appropriate for thin plates, the in-plane constitutive relation reads
\begin{align}
\sigma_{\alpha\beta} = \frac{E}{1-\nu^2}\!\left[(1-\nu)\,\varepsilon_{\alpha\beta} + \nu\,\varepsilon_{\gamma\gamma}\,\delta_{\alpha\beta}\right], \quad \alpha,\beta = 1,2.
\label{eq:plane_stress}
\end{align}
The bending-moment tensor is obtained by integrating the first moment of the in-plane stress through the thickness:
\begin{align}
M_{\alpha\beta} = \int_{-t_w/2}^{t_w/2} X_3\,\sigma_{\alpha\beta}\,dX_3 = D\!\left[(1-\nu)\,\kappa_{\alpha\beta} + \nu\,\kappa_{\gamma\gamma}\,\delta_{\alpha\beta}\right],
\label{eq:moment_tensor}
\end{align}
where
\begin{align}
D = \frac{Et_w^3}{12(1-\nu^2)}
\label{eq:flexural_rigidity}
\end{align}
is the flexural rigidity of the plate.

\noindent\textit{Equilibrium.}\; Integration of the three-dimensional balance of linear momentum \eqref{eq:blm} through the thickness, together with the traction boundary conditions on $\p\Omega^u_{top}$ and $\p\Omega^l_{top}$, yields the classical Kirchhoff--Love plate equation (see, e.g., \cite{shrimali2021remarkable} and references therein):
\begin{align}
D\,\Delta^2 w - q_{\texttt{ext}} + p - f_{\texttt{S}} - f^{\texttt{C}}(w) - f_{\texttt{B}}(w) - f_{\texttt{C}}(w) = 0 \qquad \forall\, \bar{\bfx} \in \omega,\; t \in (0,T],
\label{eq:KL_plate}
\end{align}
where $\Delta^2 \equiv \Delta(\Delta)$ is the biharmonic operator and:
\begin{itemize}
\item $q_{\texttt{ext}}(\bar{\bfx},t)$ is the resultant transverse external loading per unit area of the mid-surface, obtained by through-thickness integration of the body force :
\begin{align}
q_{\texttt{ext}}(\bar{\bfx},t) = \int_{-t_w/2}^{t_w/2} b_3\,dX_3,
\label{eq:q_ext}
\end{align}
(Note $b_3 := \mathbf{b}\cdot\mathbf{e}_3 = \rho g$ denotes the transverse ($X_3$) component of the body-force density $\mathbf{b}$.)
\item $p(\bar{\bfx},t)$ is the air pressure in the gap,
\item $f_{\texttt{B}}(w)$ is the bonding traction per unit area, and
\item $f^{\texttt{C}}(w)$ and $f_{\texttt{C}}(w)$ is the unilateral contact forces per unit area as explained in Section~\ref{subsec:contacts}.
\end{itemize}
Equation~\eqref{eq:KL_plate} is a fourth-order PDE in $w$ that inherits, through the through-thickness integration, all body and surface forces from the original three-dimensional governing equation~\eqref{eq:blm}. Note that, in these expressions, the boldface notation for the forces in \eqref{eq:blm} has been omitted to emphasize that the corresponding terms in the plate equation \eqref{eq:KL_plate} are scalar transverse components of the forces.

In the same thin-wafer limit, the Reynolds equation \eqref{eq:reynolds_3d} reduces to a two-dimensional equation on the mid-surface $\omega$. Introducing the gap
\begin{align}
h(\bar{\bfx},t) = h_0 + w(\bar{\bfx},t),
\label{eq:gap}
\end{align}
with $h_0$ denoting the initial uniform separation, the two-dimensional pressure equation is
\begin{align}
12\mu\,\frac{\p h}{\p t} - \nabla\cdot\!\left(h^3\,\nabla p\right) = 0
\qquad \forall\, \bar{\bfx} \in \omega,\; t \in (0,T].
\label{eq:reynolds_2d}
\end{align}

The reduced two-dimensional coupled FSI problem is therefore stated abstractly as:
\begin{subequations}\label{eq:2d_system}
\begin{align}
\widetilde{\mathcal{G}}_1\!\left(w(\bar{\bfx},t),\,p(\bar{\bfx},t)\right) &\;\equiv\; D\,\Delta^2 w - q_{\texttt{ext}}  + p - f_{\texttt{S}} -  f^{\texttt{C}}(w) - f_{\texttt{B}}(w) - f_{\texttt{C}}(w) = 0
&& \forall\, \bar{\bfx} \in \omega,\; t \in (0,T],
\label{eq:G1_tilde} \\[6pt]
\widetilde{\mathcal{G}}_2\!\left(p(\bar{\bfx},t),\,w(\bar{\bfx},t)\right) &\;\equiv\; 12\mu\,\frac{\p (h_0 + w)}{\p t} - \nabla\cdot\!\left((h_0 + w)^3\,\nabla p\right) = 0
&& \forall\, \bar{\bfx} \in \omega,\; t \in (0,T].
\label{eq:G2_tilde}
\end{align}
\end{subequations}

\subsection{Boundary and initial conditions}\label{subsec:bc_ic}

The coupled system \eqref{eq:G1_tilde}--\eqref{eq:G2_tilde} is supplemented with the following boundary and initial conditions for uniqueness of solutions. \rev{The boundary conditions described below apply to the top wafer, which is the only deformable body in the model. The bottom wafer is treated as rigid and fixed to the bottom chuck; its upper surface defines the reference plane $X_3 = -t_w - h_0$ against which contact is enforced.}

\noindent\textit{Mechanical boundary conditions for the plate.}\; On the outer edge $\Gamma_R = \{\bar{\bfx} \in \mathbb{R}^2 : X_1^2 + X_2^2 = R^2\}$:
\begin{subequations}\label{eq:plate_bcs}
\begin{align}
w(\bar{\bfx},t) &= w_{\Gamma}(\bar{\bfx},t) && \forall\, \bar{\bfx} \in \Gamma_R,\; t \in (0,T], \label{eq:w_bc_edge} \\
\frac{\p w}{\p n}(\bar{\bfx},t) &= 0 && \forall\, \bar{\bfx} \in \Gamma_R,\; t \in (0,T], \label{eq:w_bc_slope}
\end{align}
\end{subequations}
where $w_{\Gamma}$ is the prescribed displacement at the wafer edge (e.g., from the top chuck constraint or the release of the top wafer from the top chuck) and a zero-slope condition is imposed for a clamped boundary.

On any symmetry boundaries $\Gamma_s$ (exploited in the quarter-wafer model):
\begin{align}
\frac{\p w}{\p n}(\bar{\bfx},t) = 0, \qquad M_{nn}(w) = 0 \qquad \forall\, \bar{\bfx} \in \Gamma_s,\; t \in (0,T].
\label{eq:w_bc_sym}
\end{align}

\noindent\textit{Pressure boundary and initial conditions.}\; Let $\Gamma_{\texttt{vent}} \subseteq \p\omega$ denote the portion of the mid-surface boundary where the air gap between the wafers is connected to the ambient environment (i.e., the air pressure is prescribed). In the fully vented configuration considered here, we take $\Gamma_{\texttt{vent}}=\Gamma_R$.
\begin{subequations}\label{eq:pressure_bcs}
\begin{align}
p(\bar{\bfx},t) &= p_{\texttt{atm}} && \forall\, \bar{\bfx} \in \Gamma_{\texttt{vent}},\; t \in (0,T],
\label{eq:p_bc} \\
p(\bar{\bfx},0) &= p_0(\bar{\bfx}) && \forall\, \bar{\bfx} \in \omega,
\label{eq:p_ic}
\end{align}
\end{subequations}
where $p_{\texttt{atm}}$ is the ambient (atmospheric) pressure and $p_0$ is the initial pressure distribution (typically $p_0 = p_{\texttt{atm}}$).

\noindent\textit{Initial condition for the deflection:}
\begin{align}
w(\bar{\bfx},0) = 0 \qquad \forall\, \bar{\bfx} \in \omega.
\label{eq:w_ic}
\end{align}

\subsection{Summary of the coupled problem}\label{subsec:coupled}

For convenience, the complete coupled plate--air FSI problem is collected here. Find the transverse deflection $w(\bar{\bfx},t)$ and air pressure $p(\bar{\bfx},t)$ on the mid-surface domain $\omega$ for $t\in(0,T]$ satisfying:
\begin{subequations}\label{eq:coupled_system}
\begin{align}
\widetilde{\mathcal{G}}_1\!\left(w(\bar{\bfx},t),\,p(\bar{\bfx},t)\right) &\;\equiv\; D\,\Delta^2 w - q_{\texttt{ext}} + p - f_{\texttt{S}} - f^{\texttt{C}}(w) - f_{\texttt{B}}(w) - f_{\texttt{C}}(w) = 0
&& \forall\, \bar{\bfx} \in \omega,\; t \in (0,T],
\label{eq:coupled_plate} \\[4pt]
\widetilde{\mathcal{G}}_2\!\left(p(\bar{\bfx},t),\,w(\bar{\bfx},t)\right) &\;\equiv\; 12\mu\,\frac{\p h}{\p t} - \nabla\cdot\!\left(h^3\,\nabla p\right) = 0
&& \forall\, \bar{\bfx} \in \omega,\; t \in (0,T],
\label{eq:coupled_reynolds} \\[4pt]
h(\bar{\bfx},t) &= h_0 + w(\bar{\bfx},t)
&& \forall\, \bar{\bfx} \in \omega,\; t \in (0,T],
\label{eq:coupled_gap}
\end{align}
\end{subequations}
with the boundary conditions \eqref{eq:plate_bcs}--\eqref{eq:w_bc_sym} for $w$, the pressure conditions \eqref{eq:pressure_bcs}, and the initial conditions \eqref{eq:w_ic} and \eqref{eq:p_ic}. The coupling is two-way: the deflection $w$ modifies the gap $h$ in $\widetilde{\mathcal{G}}_2$ and thereby the air-flow resistance, while the air pressure $p$ enters the plate equation $\widetilde{\mathcal{G}}_1$ as a distributed transverse load.

\subsection{The Contacts}\label{subsec:contacts}
We model contact with the rigid chucks/wafers via (penalty-type) tractions expressed in terms of the transverse deflection $w$.\\

\noindent\textit{Top-chuck contact.}\; Let $k^{\texttt{C}}_{\texttt{top}}>0$ denote a penalty stiffness associated with the rigid top chuck. The corresponding contact traction $f^{\texttt{C}}$ is defined by
\begin{align}
    f^{\texttt{C}}(\bar{\bfx},t) \equiv f^{\texttt{C}}(w(\bar{\bfx},t)) = \begin{cases}
    k^{\texttt{C}}_{\texttt{top}}\,w(\bar{\bfx},t), & \text{if } w(\bar{\bfx},t)>0,\\
    0, & \text{otherwise}.
    \end{cases}
    \label{eq:top_chuck_contact}
\end{align}
This traction penalizes (and thereby prevents) deformation states with $w(\bar{\bfx},t)>0$ that would violate the geometric constraint imposed by the rigid top chuck.\\

\noindent\textit{Bottom-wafer contact.}\; We also introduce a unilateral contact traction $f_{\texttt{C}}$ to prevent interpenetration of the top wafer into the bottom wafer. In general, $f_{\texttt{C}}$ may be defined by an analogous penalty law activated at the closure of the gap. However, as shown in Section~\ref{subsec:results_contact}, the Reynold's pressure $p$ is sufficient to enforce the non-penetration constraint in the bonded regions and therefore provides an effective contact penalty. Accordingly, throughout the remainder of the paper we set
\begin{align}
    f_{\texttt{C}}(w) = 0.
    \label{eq:bottom_contact_suppressed}
\end{align}

\rev{\noindent\textit{Bottom wafer--chuck contact.}\; Contact between the bottom wafer and the bottom chuck is implicitly enforced through the rigid-wafer assumption: the bottom wafer remains stationary throughout the simulation, representing perfect adhesion to the chuck.}

\section{FEniCSx Implementation}\label{sec:fenicsx}

This section details the finite element discretization and solution of the coupled plate--air system \eqref{eq:coupled_system} using FEniCSx \cite{baratta2023dolfinx}. Exploiting the four-fold symmetry of the circular wafer, we solve on a quarter-domain $\omega_q = \{\bar{\bfx}\in\omega : X_1\ge 0,\; X_2\ge 0\}$ with appropriate symmetry conditions. The plate equation $\widetilde{\mathcal{G}}_1=0$ is discretized with a $C^0$ interior-penalty (C0IP) Kirchhoff--Love formulation \cite{gustafsson2020nitsche}, and the Reynolds equation $\widetilde{\mathcal{G}}_2=0$ is discretized with standard continuous Galerkin finite elements. Both fields are solved monolithically.

\subsection{Recap of the strong-form equations}\label{subsec:strong_recap}

For convenience we restate the coupled system \eqref{eq:coupled_system}. Find $w(\bar{\bfx},t)$ and $p(\bar{\bfx},t)$ satisfying:

\noindent\textit{Plate equation} (from \eqref{eq:coupled_plate}):
\begin{align}
D\,\Delta^2 w(\bar{\bfx},t) - q_{\texttt{ext}}(\bar{\bfx},t) + p(\bar{\bfx},t) - f_{\texttt{S}} - f^{\texttt{C}}(w) - f_{\texttt{B}}(w) - f_{\texttt{C}}(w) = 0
\qquad \forall\, \bar{\bfx}\in\omega_q,\; t\in(0,T],
\label{eq:fe_plate_strong}
\end{align}
where $D$ is the flexural rigidity \eqref{eq:flexural_rigidity} and the bending-moment tensor $M_{\alpha\beta}$ is given by \eqref{eq:moment_tensor}. In operator form, $D\,\Delta^2 w = M_{\alpha\beta,\alpha\beta}$, i.e.,
\begin{align}
\bfM(w) = D\!\left[(1-\nu)\,\bm{\kappa}(w) + \nu\,\mathrm{tr}(\bm{\kappa}(w))\,\bfI\right], \qquad \bm{\kappa}(w)=\nabla^2 w.
\label{eq:fe_moment}
\end{align}

\noindent\textit{Reynolds equation} (from \eqref{eq:coupled_reynolds}--\eqref{eq:coupled_gap}):
\begin{align}
12\mu\,\frac{\p h}{\p t}(\bar{\bfx},t) - \nabla\cdot\!\left(\hat{h}(\bar{\bfx},t)^3\,\nabla p(\bar{\bfx},t)\right) = 0
\qquad \forall\, \bar{\bfx}\in\omega_q,\; t\in(0,T],
\label{eq:fe_reynolds_strong}
\end{align}
with $\hat{h}(\bar{\bfx},t) = h_0 + (1-\beta)w(\bar{\bfx},t)$. \textit{Here, $0<\beta\le 0.0075$ is a small dimensionless regularization parameter; this choice is necessary to prevent $p$ from becoming unphysical as the gap closes, $w\to -h_0$.}

\rev{The regularization parameter $\beta$ maintains a minimum effective gap $\hat{h}_{\min} = \beta h_0$ when full contact occurs ($w \to -h_0$). This regularization has both physical and numerical justifications:
\begin{itemize}
    \item \textit{Physical basis:} Even for highly polished silicon wafers, sub-micron surface roughness prevents perfect molecular contact. The root-mean-square roughness of CMP-polished wafers is typically 0.1--0.5~nm~\cite{teichert1995comparison}, but asperity peaks can extend to several nanometers. The regularized minimum gap ($\beta h_0 \approx 0.75\,\mu$m for $h_0 = 100\,\mu$m and $\beta = 0.0075$) is conservative relative to this physical roughness scale.
    \item \textit{Numerical justification:} The $h^3$ term in the Reynolds equation causes numerical ill-conditioning as $h \to 0$. The regularization ensures that the discrete system remains well-posed and that the Newton solver converges reliably.
    \item \textit{Continuum validity:} As discussed in Section~\ref{sec:theory} (condition~(iv)), the Knudsen number $\text{Kn} = \lambda/h$ governs the transition between continuum and rarefied gas flow regimes~\cite{burgdorfer1959influence, fukui1988analysis}. At the regularized minimum gap $\hat{h}_{\min} = \beta h_0 = 0.75\,\mu$m ($\beta = 0.0075$, $h_0 = 100\,\mu$m), the Knudsen number at atmospheric pressure is $\text{Kn} \approx 68\,\text{nm}/0.75\,\mu\text{m} \approx 0.09$; however, the simultaneously rising pressure near the bond front reduces $\lambda$ (since $\lambda \propto 1/P$), keeping the flow within the slip-flow regime ($\text{Kn} < 0.1$) where the Reynolds equation remains applicable~\cite{burgdorfer1959influence}.
    \item \textit{Asymptotic validity:} The mathematical validity of the Reynolds equation as a rigorous limit of the Navier--Stokes equations is established through asymptotic analysis under small aspect ratios and geometric perturbations~\cite{bayada1986transition}. Perturbing the film thickness by the small scaling factor $\beta = 0.0075$ confines the introduced error to $\mathcal{O}(\beta^2)$, well within the accepted tolerances of the thin-film approximation.
\end{itemize}
A sensitivity analysis demonstrating that the results are insensitive to $\beta$ in the range $[0.0025, 0.01]$ is presented in Appendix~\ref{app:beta_sensitivity}.}\\

\noindent\textit{Boundary conditions.}\; On $\omega_q$ the boundary $\p\omega_q$ is decomposed into:
\begin{itemize}
\item \textit{Symmetry edges} $\Gamma_x = \{\bar{\bfx}\in\p\omega_q : X_1=0\}$ and $\Gamma_y = \{\bar{\bfx}\in\p\omega_q : X_2=0\}$, where the slope vanishes, $\p_n w = 0$, and $M_{nn}(w) = 0$ (cf.\ \eqref{eq:w_bc_sym}).
\item \textit{Outer arc} $\Gamma_R = \{\bar{\bfx}\in\p\omega_q : X_1^2+X_2^2=R^2\}$, where displacement Dirichlet conditions $w = w_{\Gamma}$ (cf.\ \eqref{eq:w_bc_edge}) and the pressure venting condition $p_{\texttt{air}} = p_\infty$ (cf.\ \eqref{eq:p_bc}) are imposed.
\end{itemize}

\begin{figure}[htbp]
\centering
\begin{subfigure}[b]{0.49\textwidth}
\centering
\begin{tikzpicture}[scale=2.6, >=stealth]
  \fill[blue!6] (0,0) -- (1.6,0) arc[start angle=0, end angle=90, radius=1.6] -- cycle;

  \draw[red!80!black, line width=1.6pt, dashed] (0,0) -- (1.6,0);
  \draw[red!80!black, line width=1.6pt, dashed] (0,0) -- (0,1.6);
  \draw[red!80!black, line width=1.6pt] (1.6,0) arc[start angle=0, end angle=90, radius=1.6];

  \draw[->, gray, line width=0.6pt] (-0.35,0) -- (2.05,0) node[right, black]{$\bar{x}_1$};
  \draw[->, gray, line width=0.6pt] (0,-0.35) -- (0,2.05) node[above, black]{$\bar{x}_2$};

  \fill[black] (0,0) circle(0.6pt);
  \node[font=\scriptsize, gray, below left=2pt] at (0,0) {$O$};
  \draw[gray, line width=0.4pt] (0.08,0) -- (0.08,0.08) -- (0,0.08);

  \node at (0.7,0.85) {\large$\omega_q$};

  \fill[blue!25, opacity=0.5] (0,0) -- (0.35,0) arc[start angle=0, end angle=90, radius=0.35] -- cycle;
  \draw[blue!60!black, line width=0.8pt, densely dotted] (0.35,0) arc[start angle=0, end angle=90, radius=0.35];
  \node[blue!70!black, font=\footnotesize] at (0.12,0.12) {$\omega_S$};
  \draw[->, blue!60!black, line width=0.5pt] (0.155,0.18) -- ({0.3*cos(55)+0.02},{0.3*sin(55)+0.02});
  \node[blue!70!black, font=\scriptsize] at (0.4,0.34) {(striker)};

  \node[red!80!black, above=3pt, font=\small] at (1.05,0) {$\Gamma_y\;(\partial_n w = 0)$};
  \node[red!80!black, rotate=90, anchor=south, font=\small] at (0.22,1.05) {$\Gamma_x\;(\partial_n w = 0)$};
  \node[red!80!black] at ({1.25*cos(65)},{1.55*sin(65)}) {$\Gamma_R$};
  \node[red!80!black, font=\footnotesize, align=left] at ({1.05*cos(40)},{1.05*sin(40)}) {$w = w_\Gamma$\\$p = p_\infty$};

  \draw[<->, gray, line width=0.5pt] (0,-0.42) -- (1.6,-0.42);
  \node[gray, below=1pt, font=\footnotesize] at (0.8,-0.42) {$R$};

  \draw[->, red!80!black, line width=0.8pt] (0.55,0) -- (0.55,-0.17) node[below, font=\scriptsize]{$\bfn$};
  \draw[->, red!80!black, line width=0.8pt] (0,0.65) -- (-0.17,0.65) node[left, font=\scriptsize]{$\bfn$};
  \draw[->, red!80!black, line width=0.8pt] ({1.6*cos(25)},{1.6*sin(25)}) -- ({(1.6+0.18)*cos(25)},{(1.6+0.18)*sin(25)}) node[right, font=\scriptsize]{$\bfn$};
\end{tikzpicture}
\caption{Schematic of the quarter-domain $\omega_q$.}
\label{fig:quarter_domain_schematic}
\end{subfigure}
\hfill
\begin{subfigure}[b]{0.48\textwidth}
\centering
\includegraphics[width=\linewidth]{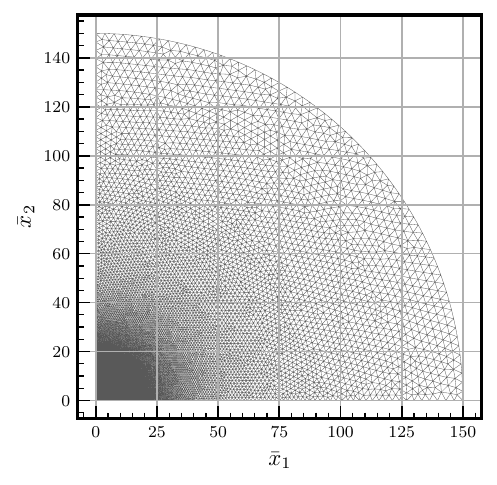}
\caption{Finite-element triangulation of $\omega_q$ ($42{,}449$ triangular elements).}
\label{fig:quarter_domain_mesh}
\end{subfigure}
\caption{Quarter-domain $\omega_q$ exploiting four-fold symmetry of the circular wafer mid-surface $\omega$.
(a)~Red boundaries: dashed lines denote symmetry edges $\Gamma_x,\Gamma_y$ (zero normal slope); solid arc denotes the outer boundary $\Gamma_R$ (Dirichlet conditions). The striker region $\omega_S$ is shown near the origin.
(b)~The corresponding finite-element mesh ($42{,}449$ triangular elements) used in the \texttt{DOLFINx} computation.}
\label{fig:quarter_domain}
\end{figure}


\subsection{Time discretization}\label{subsec:time_disc}

The Reynolds equation \eqref{eq:fe_reynolds_strong} is the only equation containing a time derivative. We discretize it with an implicit (backward) Euler scheme. Partitioning $(0,T]$ into time steps $0 = t_0 < t_1 < \cdots < t_N = T$ with $\Delta t_n = t_{n+1}-t_n$, and denoting $w^n \approx w(\cdot,t_n)$, $p^{n+1}\approx p_{\texttt{air}}(\cdot,t_{n+1})$, the semi-discrete system at step $n+1$ reads:
\begin{subequations}\label{eq:fe_semidiscrete}
\begin{align}
D\,\Delta^2 w^{n+1} - q_{\texttt{ext}}^{n+1} + p^{n+1}_{\texttt{air}} - f_{\texttt{S}}(w^{n+1}) - f^{\texttt{C}}(w^{n+1}) - f_{\texttt{B}}(w^{n+1}) - f_{\texttt{C}}(w^{n+1}) &= 0
&& \forall\, \bar{\bfx}\in\omega_q,
\label{eq:fe_semi_plate} \\[4pt]
\frac{12\mu}{\Delta t_n}\!\left(w^{n+1}-w^n\right) - \nabla\cdot\!\left(\hat{h}(w^{n+1})^3\,\nabla p^{n+1}\right) &= 0
&& \forall\, \bar{\bfx}\in\omega_q.
\label{eq:fe_semi_reynolds}
\end{align}
\end{subequations}

\subsection{Weak formulation}\label{subsec:weak_form}

Let $V = H^2(\omega_q)$ and $Q = H^1(\omega_q)$ denote the trial/test spaces for the deflection and pressure fields, respectively, and let $v\in V$, $q\in Q$ be test functions. Multiplying \eqref{eq:fe_semi_plate} by $v$, integrating by parts twice, and multiplying \eqref{eq:fe_semi_reynolds} by $q$ with a single integration by parts, yields the following weak residuals.\\

\noindent\textit{Plate residual.}\; Starting from $\int_{\omega_q} M_{\alpha\beta,\alpha\beta}\,v\,d\omega = 0$, two integrations by parts give
\begin{align}
\mathcal{R}_w(w^{n+1},p^{n+1};\,v) &=
\int_{\omega_q} \bfM(w^{n+1}):\nabla^2 v\,d\omega
+ \int_{\omega_q} \left(p^{n+1} - q_{\texttt{ext}}^{n+1}\right) v\,d\omega
\nonumber\\
&\quad + \int_{\omega_q} \left[-f_{\texttt{S}}(w^{n+1})-f_{\texttt{B}}(w^{n+1}) - f^{\texttt{C}}(w^{n+1}) + f_{\texttt{C}}(w^{n+1})\,w^{n+1}\right] v\,d\omega
\label{eq:fe_Rw_cont}
\end{align}
where all boundary terms vanish on $\Gamma_R$ (essential BC) and on $\Gamma_x\cup\Gamma_y$ (symmetry, $M_{nn}=0$, $\p_n w = 0$).\\

\noindent\textit{Pressure residual.}\; Multiplying \eqref{eq:fe_semi_reynolds} by $q$ and integrating by parts:
\begin{align}
\mathcal{R}_p(w^{n+1},p^{n+1};\,q) &=
\int_{\omega_q} 12\mu\left(w^{n+1}-w^n\right) q\,d\omega
+ \Delta t_n\!\int_{\omega_q} \hat{h}(w^{n+1})^3\,\nabla p^{n+1}\cdot\nabla q\,d\omega,
\label{eq:fe_Rp}
\end{align}
where the boundary flux term vanishes on $\Gamma_R$ ($p^{n+1}=p_\infty$, Dirichlet) and on the symmetry edges (zero normal flux by symmetry).

\subsection{\texorpdfstring{$C^0$}{C0} interior-penalty spatial discretization}\label{subsec:c0ip}

Since the plate weak form \eqref{eq:fe_Rw_cont} involves second derivatives of $w$, it requires $H^2$-conforming elements. Instead, we use the $C^0$ interior-penalty (C0IP) method \cite{gustafsson2020nitsche}, which employs standard $C^0$-continuous Lagrange elements and enforces inter-element $C^1$-continuity weakly through penalty terms on interior facets.

\noindent\textit{Finite element spaces.}\; Let $\mathcal{T}_h$ be a conforming triangulation of $\omega_q$ and define the quadratic Lagrange spaces
\begin{align}
V_h &= \accol{\phi_h \in C^0(\overline{\omega}_q) : \phi_h\big|_K \in \mathbb{P}_2(K)\;\;\forall\, K\in\mathcal{T}_h},
\label{eq:fe_Vh}\\
Q_h &= \accol{\psi_h \in C^0(\overline{\omega}_q) : \psi_h\big|_K \in \mathbb{P}_2(K)\;\;\forall\, K\in\mathcal{T}_h},
\label{eq:fe_Qh}
\end{align}
and the mixed space $W_h = V_h \times Q_h$.

\noindent\textit{Interior-penalty terms.}\; Let $\mathcal{F}_h^i$ denote the set of interior facets of $\mathcal{T}_h$, and for any facet $e\in\mathcal{F}_h^i$ let $\bfn_e$ be a fixed unit normal, $h_e$ the facet diameter, and define the average $\{\!\{\cdot\}\!\}$ and jump $\croch{\cdot}$ operators. The normal bending moment and normal slope on a facet are
\begin{align}
M_{nn}(w_h) = \bfn_e\cdot\bfM(w_h)\,\bfn_e, \qquad
\croch{\p_n w_h} = \croch{\nabla w_h \cdot \bfn_e}.
\label{eq:fe_Mnn_jump}
\end{align}

The C0IP bending residual, replacing the continuous form \eqref{eq:fe_Rw_cont}, reads
\begin{align}
\mathcal{R}_w^h(w_h^{n+1},p_h^{n+1};\,v_h)
&= \int_{\omega_q} \bfM(w_h^{n+1}):\nabla^2 v_h\,d\omega
\nonumber\\
&\quad - \int_{\mathcal{F}_h^i}\!\{\!\{M_{nn}(w_h^{n+1})\}\!\}\,\croch{\p_n v_h}\,d\Gamma
- \int_{\mathcal{F}_h^i}\!\{\!\{M_{nn}(v_h)\}\!\}\,\croch{\p_n w_h^{n+1}}\,d\Gamma
\nonumber\\
&\quad + \int_{\mathcal{F}_h^i} \frac{\alpha_{\mathrm{ip}}\,D}{h_e}\,\croch{\p_n w_h^{n+1}}\,\croch{\p_n v_h}\,d\Gamma + \int_{\omega_q}\!\left(p_h^{n+1} - q_{\texttt{ext}}^{n+1}\right) v_h\,d\omega
\nonumber\\
&\quad + \int_{\omega_q}\!\left[-f_{\texttt{S}}(w_h^{n+1})-f_{\texttt{B}}(w_h^{n+1}) - f^{\texttt{C}}(w_h^{n+1}) + f_{\texttt{C}}(w_h^{n+1})\,w_h^{n+1}\right] v_h\,d\omega,
\label{eq:fe_Rw_c0ip}
\end{align}
where $\alpha_{\mathrm{ip}}>0$ is the interior-penalty parameter.

On the symmetry boundaries $\Gamma_s\in\{\Gamma_x,\Gamma_y\}$, the condition $\p_n w = 0$ (cf.\ \eqref{eq:w_bc_sym}) is enforced weakly via Nitsche-type terms consistent with the C0IP formulation:
\begin{align}
\mathcal{R}_{\mathrm{sym}}^h(w_h^{n+1};\,v_h) = \sum_{\Gamma_s\in\{\Gamma_x,\Gamma_y\}}\!\left[
-\int_{\Gamma_s}\!M_{nn}(w_h^{n+1})\,\p_n v_h\,d\Gamma
-\int_{\Gamma_s}\!M_{nn}(v_h)\,\p_n w_h^{n+1}\,d\Gamma
+\int_{\Gamma_s}\!\frac{\alpha_{\mathrm{ip}}\,D}{h_e}\,\p_n w_h^{n+1}\,\p_n v_h\,d\Gamma
\right].
\label{eq:fe_Rsym}
\end{align}

The discrete pressure residual is identical to the continuous form \eqref{eq:fe_Rp} restricted to $Q_h$:
\begin{align}
\mathcal{R}_p^h(w_h^{n+1},p_h^{n+1};\,q_h) =
\int_{\omega_q} 12\mu\left(w_h^{n+1}-w_h^n\right) q_h\,d\omega
+ \Delta t_n\!\int_{\omega_q} \hat{h}(w_h^{n+1})^3\,\nabla p_h^{n+1}\cdot\nabla q_h\,d\omega.
\label{eq:fe_Rp_h}
\end{align}

\subsection{Monolithic nonlinear system and Newton solver}\label{subsec:newton}

The fully discrete monolithic residual is assembled as
\begin{align}
\mathcal{R}^h(w_h^{n+1},p_h^{n+1};\,v_h,q_h) =
\mathcal{R}_w^h(w_h^{n+1},p_h^{n+1};\,v_h)
+ \mathcal{R}_{\mathrm{sym}}^h(w_h^{n+1};\,v_h)
+ \mathcal{R}_p^h(w_h^{n+1},p_h^{n+1};\,q_h).
\label{eq:fe_residual}
\end{align}

At each time step $t_{n+1}$, we seek $(w_h^{n+1},\,p_h^{n+1})\in W_h$ such that
\begin{align}
\mathcal{R}^h(w_h^{n+1},p_h^{n+1};\,v_h,q_h)=0
\qquad \forall\,(v_h,q_h)\in W_h^0,
\label{eq:fe_discrete_problem}
\end{align}
where $W_h^0\subset W_h$ incorporates homogeneous essential boundary conditions ($v_h = 0$ on $\Gamma_R$, $q_h = 0$ on $\Gamma_{\texttt{vent}}$). The essential (Dirichlet) conditions enforced are:
\begin{itemize}
\item $p_h^{n+1} = 0$ on the vented outer boundary $\Gamma_R$ (gauge pressure),
\item $w_h^{n+1} = w_{\mathrm{edge}}(t_{n+1})$ on $\Gamma_R$,
\item $w_h^{n+1} = w_{\mathrm{center}}(t_{n+1})$ at the striker/pin region.
\end{itemize}

Due to the nonlinear loading terms $f_{\texttt{B}}(w)$, $f_c(w)$, $f_{\mathrm{tc}}(w)$ and the cubic gap dependence $\hat{h}(w)^3$ in the Reynolds equation, the discrete problem \eqref{eq:fe_discrete_problem} is nonlinear and is solved by Newton's method. Denoting the algebraic unknown vector $\bfU = (\bfw_h,\,\mathbf{p}_h)^\top$ and the algebraic residual vector $\bfR(\bfU)$, the Newton iteration reads
\begin{align}
\mathbf{J}(\bfU^{(k)})\,\Delta\bfU^{(k)} = -\bfR(\bfU^{(k)}),
\qquad
\bfU^{(k+1)} = \bfU^{(k)} + \Delta\bfU^{(k)},
\label{eq:fe_newton}
\end{align}
where the Jacobian $\mathbf{J} = \p\bfR/\p\bfU$ is computed automatically via UFL's \cite{UFL} symbolic differentiation capabilities in FEniCSx.

The complete solution procedure is summarized in Algorithm~\ref{alg:solve}.

\begin{algorithm}[tbp]
\caption{Monolithic coupled wafer-wafer bonding simulation in \texttt{FEniCSx}}\label{alg:solve}
\begin{algorithmic}[1]

\STATE \texttt{// ---- Mesh, FE Spaces \& Measures ----}
\STATE Load Gmsh model via \texttt{gmsh.model\_to\_mesh} $\to$ mesh $\mathcal{T}_h$;\; build connectivity
\STATE $h_{\mathrm{ip}} \leftarrow \sqrt{|\omega_q|/N_{\mathrm{cells}}}$ \hfill \textit{IP penalty scale}
\STATE $W_h \leftarrow$ \texttt{functionspace(msh, mixed\_element([P2, P2]))}
\STATE $(w_h,p_h),\,(v_h,q_h),\,(w_h^n,p_h^n) \leftarrow$ \texttt{split} of trial, test, prev.\ step \texttt{Function}s
\STATE \texttt{dx}, \texttt{dS} $\leftarrow$ cell/interior-facet measures;\; \texttt{ds\_sym} $\leftarrow$ tagged symmetry facets via \texttt{meshtags}

\medskip
\STATE \texttt{// ---- UFL Variational Forms ----}
\STATE $\triangleright$ Plate residual $\mathcal{R}_w^h$ \eqref{eq:fe_Rw_c0ip}: \texttt{F1} $\leftarrow$ \texttt{inner(M,hess)*dx} $-$ \texttt{avg(M\_nn)*jump*dS} $+$ penalty $+$ loads
\STATE $\triangleright$ Nitsche symmetry $\mathcal{R}_{\mathrm{sym}}^h$ \eqref{eq:fe_Rsym}: \texttt{F1} $\mathrel{+}=$ $\sum_{\Gamma_s}$[\texttt{-M\_nn*dn\_v} $+$ $\alpha D/h$\texttt{*dn\_w*dn\_v}]\texttt{*ds\_sym}
\STATE $\triangleright$ Pressure residual $\mathcal{R}_p^h$ \eqref{eq:fe_Rp_h}: \texttt{F2} $\leftarrow$ $12\mu$\texttt{(w-wn)*q*dx} $+$ $\Delta t$\texttt{*}\,$\hat{h}^3$\,\texttt{*dot(grad(p),grad(q))*dx}
\STATE \texttt{residual} $\leftarrow$ \texttt{F1+F2};\quad \texttt{jacobian} $\leftarrow$ \texttt{derivative(residual, w\_p, TrialFunction)} \hfill \textit{\cite{UFL}}

\medskip
\STATE \texttt{// ---- Boundary Conditions \& Solver ----}
\STATE \texttt{bcs} $\leftarrow$ [\texttt{dirichletbc}($p{=}0$, $\Gamma_R$),\; \texttt{dirichletbc}($w{=}w_{\mathrm{edge}}$, $\Gamma_R$),\; \texttt{dirichletbc}($w{=}w_{\mathrm{center}}$, $\omega_S$)]
\STATE \texttt{problem} $\leftarrow$ \texttt{NonlinearProblem(residual, w\_p, bcs, J=jacobian)};\; PETSc LU/MUMPS
\STATE Open \texttt{VTXWriter} for $w$, $p$ output (BP4)

\medskip
\STATE \texttt{// ---- Time-Stepping Loop ----}
\FOR{$n = 0,\,1,\,\ldots,\,N{-}1$}
  \STATE $t_{n+1} \leftarrow t_n + \Delta t_n$;\; update \texttt{dt.value}, \texttt{disp\_edge.value}, \texttt{disp\_center.value} from CSV
  \STATE \textbf{Solve:} \texttt{problem.solve()} \hfill \textit{assembles $\bfR$, $\mathbf{J}$; Newton iteration \eqref{eq:fe_newton}}
  \STATE \texttt{w\_p\_n.x.array[:] = w\_p.x.array[:]};\; \texttt{w\_sol.interpolate(w\_sub)}
  \STATE \texttt{w\_vtx.write($t_{n+1}$)};\; \texttt{p\_vtx.write($t_{n+1}$)} \hfill \textit{write output}
\ENDFOR

\medskip
\STATE \texttt{w\_vtx.close()};\; \texttt{p\_vtx.close()};\; save radial profiles via \texttt{np.save}
\RETURN $(w_h^N,\,p_h^N)$

\end{algorithmic}
\end{algorithm}

\section{Results}
\label{sec:results}
This section presents the numerical results for the wafer-bonding model; the solution workflow is outlined in Algorithm~\ref{alg:solve}. We first validate the implementation against experimental top-wafer displacement measurements and illustrate the resulting quarter-symmetric displacement field (Figure~\ref{fig:displacement_warped_3x3}). We then verify the force balance at the bond front by comparing the Reynolds air pressure with the computed contact reaction (Section~\ref{subsec:results_contact}). Next, we assess how the initial bond gap affects air-pressure build-up and the bond-front propagation speed (Section~\ref{sec:bond_gap_bond_speed}), and we conduct a parametric study on the role of air viscosity. Finally, we examine how the interfacial energy influences the bonding speed.

\rev{\textbf{Terminology note:} Here ``bond speed'' denotes the radial gap-closure rate $\dot r_b=dr_b/dt$, where $r_b$ is the radius from the center-connected region satisfying the bonding criterion $h\leq\delta$ (region already bonded). It is a post-processed radial propagation rate obtained from the out-of-plane Kirchhoff--Love plate solution, rather than a material-point velocity.}

\rev{Table~\ref{tab:input_parameters} summarizes the input parameters common to all simulations presented in this section. Parameters marked with ${}^\dagger$ were calibrated as described in Section~\ref{subsec:results_validation}.}

\begin{table}[htbp]
\centering
\rev{\caption{Input parameters for the wafer bonding simulations.}
\label{tab:input_parameters}
\begin{tabular}{llll}
\hline
Parameter & Symbol & Value & Source \\
\hline
Wafer radius & $R$ & $150\,\mathrm{mm}$ & Standard 12" wafer \\
Wafer thickness & $t_w$ & $0.775\,\mathrm{mm}$ & Standard 12" wafer \\
Young's modulus & $E$ & $148\,\mathrm{GPa}$ & Standard Si \\
Poisson's ratio & $\nu$ & $0.27$ & Standard Si \\
Air viscosity & $\mu$ & $1.86 \times 10^{-5}\,\mathrm{Pa\!\cdot\!s}$ & Standard (20$^\circ$C) \\
Pin (striker) radius & $R_{\mathrm{\texttt{S}}}/R_{pin}$ & $5.0\,\mathrm{mm}$ & Assumed \\
Initial bond gaps & $h_0$ & $30,\;70,\;100\,\mu\mathrm{m}$ & \cite{ip2022multi} \\
Regularization parameter & $\beta$ & $0.0075$ & Sec.~\ref{subsec:strong_recap}; App.~\ref{app:beta_sensitivity} \\
Bonding traction${}^\dagger$ & $f_{\texttt{BE}}$ & $0.03\,\mathrm{MPa}$ & Calibrated \\
Bonding threshold${}^\dagger$ & $\delta$ & $0.001\,\mathrm{mm}$ & Calibrated \\
Interfacial energy${}^\dagger$ & $\Gamma = f_{\texttt{BE}}\delta$ & $0.03\,\mathrm{J/m}^2$ & Calibrated \\
Striker pressure${}^\dagger$ & $f_{\texttt{S}}$ & $0.028\,\mathrm{MPa}$ & Calibrated \\
Displacement location & $r_{\texttt{mid-region}}$ & $75\,\mathrm{mm}$ & Assumed \\
\hline
\end{tabular}
\end{table}
}

\subsection{Validation}
\label{subsec:results_validation}
\begin{figure}[H]
    \centering
    \includegraphics[width=0.82\textwidth]{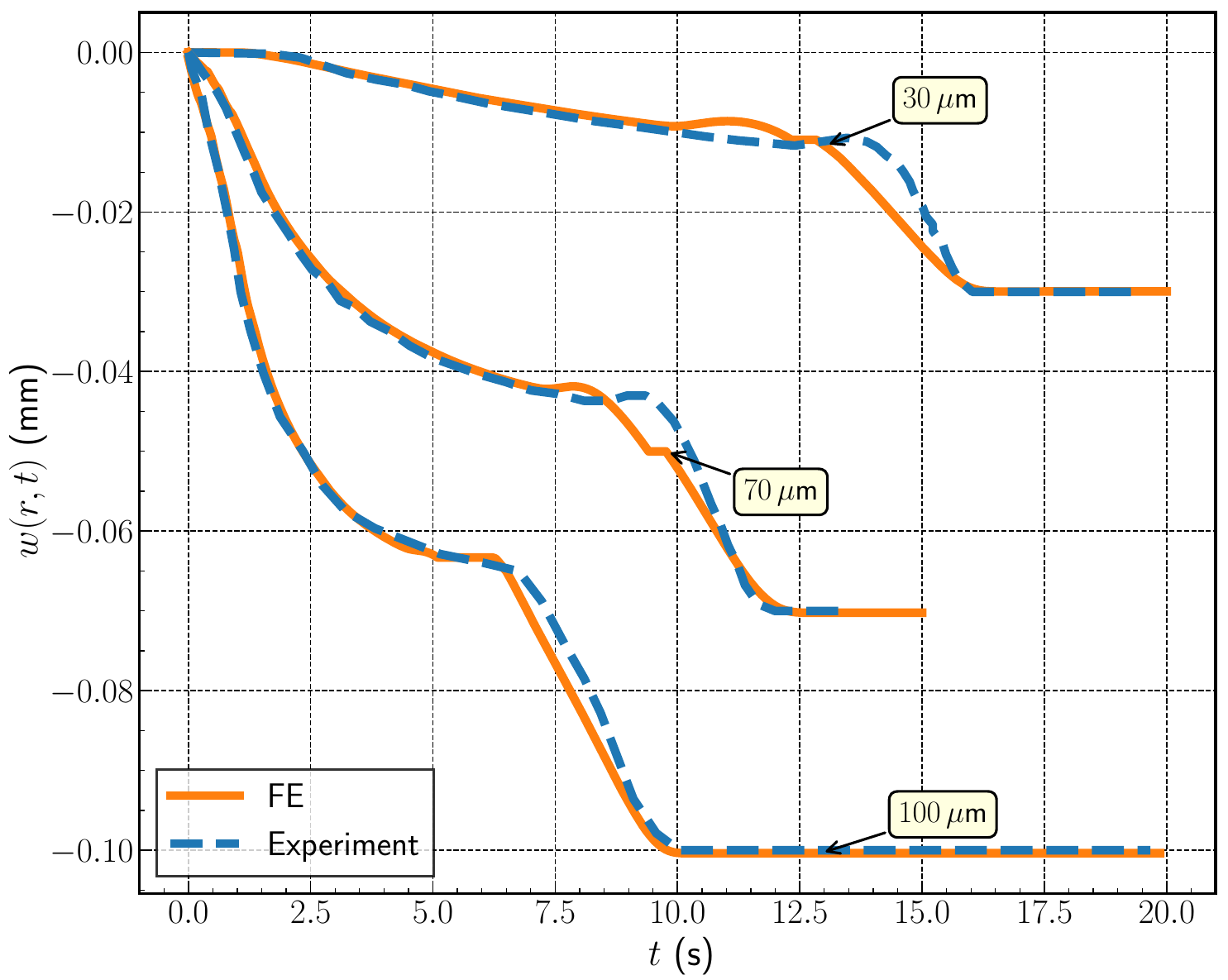}
    \caption{Top wafer displacement $w(r,t)$ versus time at a fixed radial location $r = \textrm{\textit{mid-region}}$
             for $h_0 = 30$, $70$, and $100\,\mu$m.
             Solid lines show the finite-element solution; dashed lines show the experimental measurement.}
    \label{fig:probe_displacement}
\end{figure}

To validate the implementation, we performed three simulations corresponding to the bond gaps (a) $h_0 = 30\,\mu\mathrm{m}$, (b) $h_0 = 70\,\mu\mathrm{m}$, and (c) $h_0 = 100\,\mu\mathrm{m}$. In each case, Dirichlet boundary condition $w = w_{\Gamma}$ was prescribed using the displacement history reported in Figure~3 of \cite{ip2022multi}. The resulting displacement time history at the mid-region\footnote{\rev{Note that the precise location of this \textit{mid-region} was assumed to be at 75 mm from the top wafer center. This is because the location of this region perhaps represents intellectual property of the bonder apparatus manufacturer and was missing in \cite{ip2022multi}.}} of the top wafer was then compared with the experimental measurements. Figure~\ref{fig:probe_displacement} shows that the model reproduces the measured response with good qualitative agreement and reasonable quantitative accuracy. Small discrepancies are expected because several inputs required for an exact match are not fully specified in \cite{ip2022multi} (e.g., interfacial energy and the precise probe location); these were therefore inferred through a calibration procedure in this work.

\rev{The parameters not specified in \cite{ip2022multi} are the striker pressure $f_{\texttt{S}}$, the bonding traction magnitude $f_{\texttt{BE}}$ (Eq.~\eqref{eq:BE}), the bonding activation threshold $\delta$, and the radial location of the displacement sensor. The sensor location was assumed to be $r = 75\,\mathrm{mm}$ based on the reported experimental setup. The striker pressure, the bonding traction and threshold were calibrated by minimizing the discrepancy between the simulated and measured displacement time histories at the sensor location. The calibrated values are $f_{\texttt{BE}} = 0.03\,\mathrm{MPa}$ and $\delta = 0.001\,\mathrm{mm}$ ($= 1\,\mu\mathrm{m}$). The product $\Gamma = f_{\texttt{BE}}\,\delta = 0.03\,\mathrm{J/m}^2$ corresponds to the interfacial bonding energy, which falls within the range of $20$--$100\,\mathrm{mJ/m}^2$ reported for room-temperature hydrophilic silicon wafer bonding \cite{gosele1998semiconductor,larrey2016adhesion}. Furthermore, since the wafer release condition was not reported in \cite{ip2022multi}, the measured edge displacement time history was instead prescribed as a displacement boundary condition on the edge $\Gamma_R$. The striker pressure, applied uniformly over the central region $r \le R_{\texttt{S}}$, was calibrated by matching the measured wafer-center displacement history, giving $f_{\texttt{S}} = 0.028\,\mathrm{MPa}$}.

The effect of air pressure is evident in both the experiments and the simulations. In contrast to what is expected for bonding in the absence of air (pure-plate-bending), in the presence of air, the largest initial bond gap reaches complete bonding earliest, whereas the smallest gap reaches it last. This ordering is also evident in the bonded-radius evolution discussed in Section~\ref{sec:bond_gap_bond_speed}.

\begin{figure}[t]
    \centering
    \includegraphics[width=\textwidth]{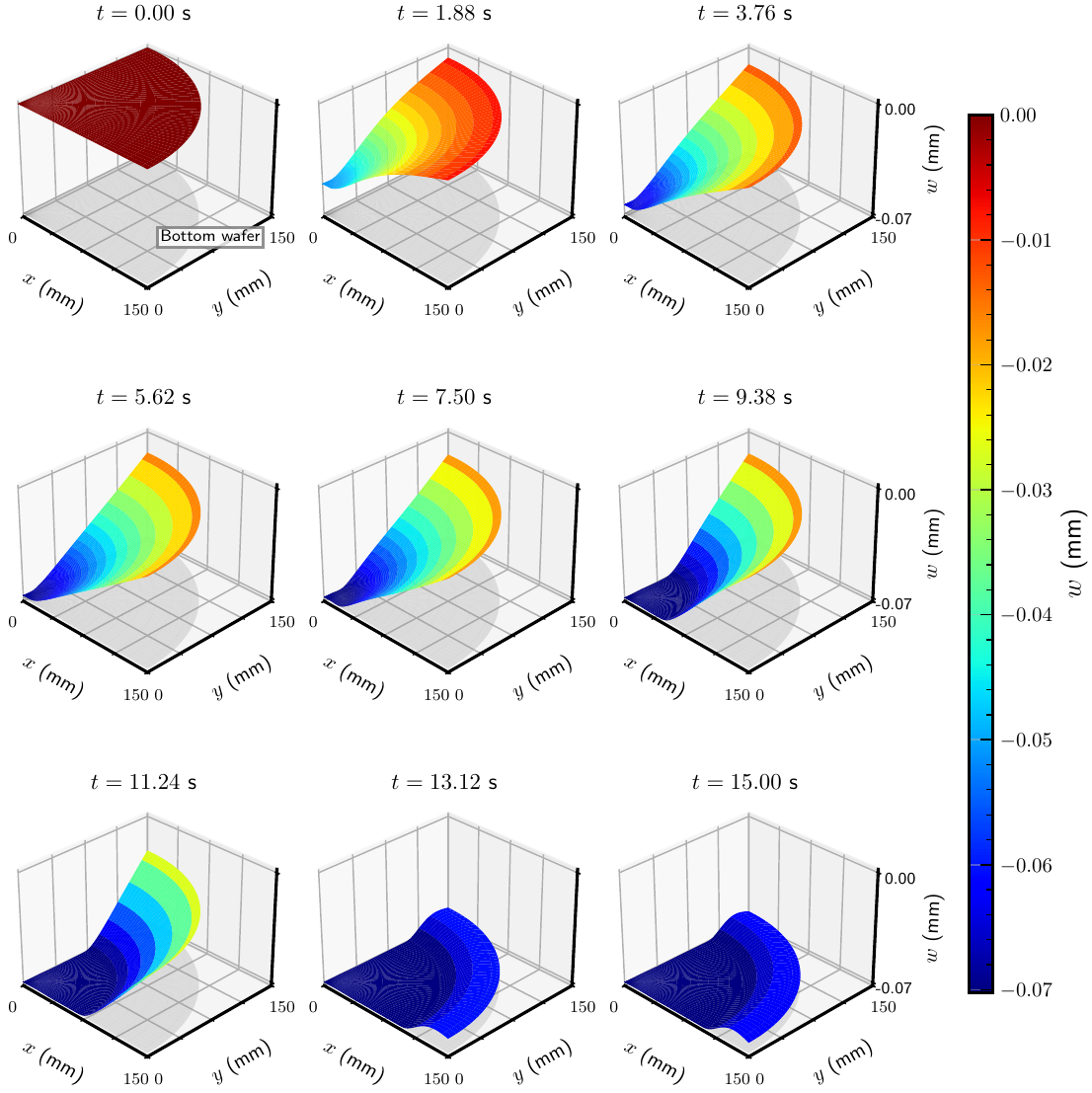}
    \caption{Snapshots of the displacement field $w$ at nine time instances during bonding for $h_0=70\,\mu\mathrm{m}$. The out-of-plane displacement is warped by a factor of 500 to enhance the visualization. The bottom wafer is shown in gray, and the color scale indicates the physical displacement of the top wafer.}
    \label{fig:displacement_warped_3x3}
\end{figure}

Snapshots of the simulated out-of-plane displacement field are shown as 2D contour plots at selected time instants in Figure~\ref{fig:displacement_warped_3x3}.

\FloatBarrier

\subsection{Reynold's pressure acting as a contact force}\label{subsec:results_contact}
\begin{figure}[htbp]
    \centering
    \includegraphics[width=\textwidth]{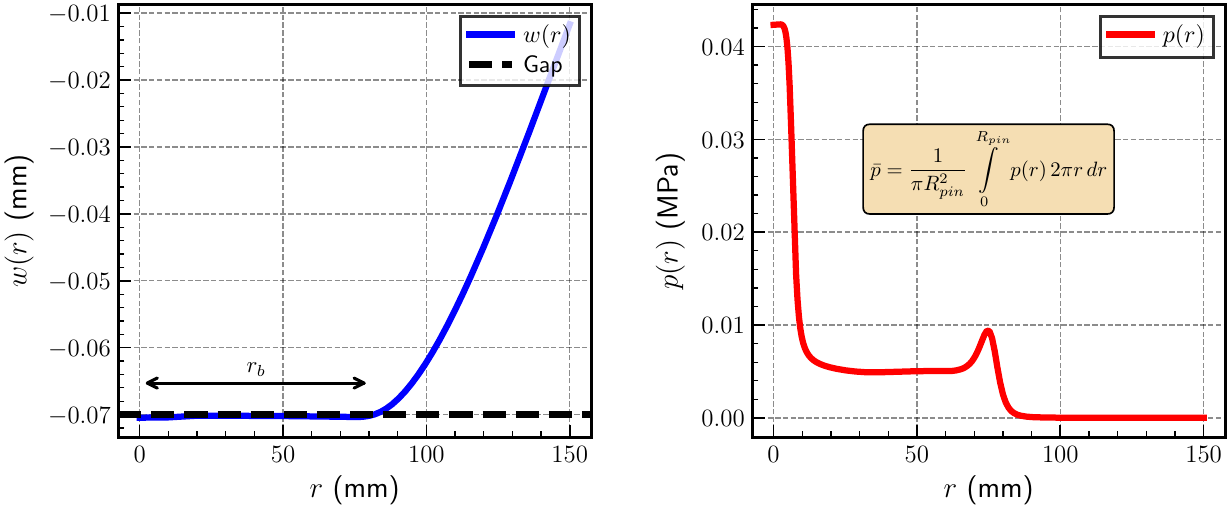}
    \caption{Displacement $w(r)$ (left) and pressure $p(r)$ (right) at the final time $t=50\,\mathrm{s}$ for $h_0=70\,\mu\mathrm{m}$. The average pressure $\bar p$ is computed by integrating the Reynolds pressure over $r\leq R_{\mathrm{S}}=R_{pin}$; the resulting average pressure is $\bar p= f_{\texttt{S}} = 0.028\,\mathrm{MPa}$.}
    \label{fig:contact_pressure_verification}
\end{figure}

As discussed in Section~\ref{subsec:contacts}, we set the bottom-wafer contact tractions to zero, $f_{\texttt{C}}(w)=0$, because the air-film pressure obtained from the Reynolds solution, $p$, is sufficient to enforce non-penetration contact. This can be seen from Eq.~\eqref{eq:fe_reynolds_strong}: the Reynolds equation is solved over the full computational domain $\omega$, including regions that are already bonded. In the bonded region the Reynolds pressure acts as a reaction pressure, i.e., an effective contact traction that prevents interpenetration of the wafers. This is because of the inherent mathematical structure of the Reynold's equation in which $h$ in Eq.~\eqref{eq:fe_reynolds_strong} cannot be negative. Figure~\ref{fig:contact_pressure_verification} confirms this interpretation by showing that the Reynolds pressure behind the bond front balances the applied striker pressure. So, at the bond front, the total Reynold's pressure $p$ is :
\begin{align}
    p(\bar{\bfx},t) = \bar{p}(\bar{\bfx},t) + p_{\texttt{air}}(\bar{\bfx},t),
    \label{eq:bond_front_pressure_balance}
\end{align}
where, $p_{\texttt{air}}$ is the actual pressure due to the air between the two wafers in the unbonded region and $\bar{p}$ is the contact pressure.

\rev{We note that the Reynolds pressure computed in the bonded region (where the effective gap approaches $\hat{h} \to \beta h_0$) does not represent physical air pressure, as the air has been expelled from bonded areas. Rather, this pressure serves as an effective contact reaction that enforces the non-penetration constraint, analogous to a penalty-regularized contact formulation. This interpretation is validated by the force balance shown in Figure~\ref{fig:contact_pressure_verification}. The regularization parameter $\beta$ (cf.\ Section~\ref{subsec:strong_recap}) is essential for this contact-reaction interpretation: by maintaining a minimum effective gap, it prevents the pressure singularity that would occur as $h \to 0$ in the unregularized Reynolds equation while allowing the Reynolds equation to return a finite, well-defined pressure in bonded regions.}

\FloatBarrier

\subsection{Effect of bond gap on bond speed}\label{sec:bond_gap_bond_speed}

\begin{figure}[htbp]
    \centering
    \begin{subfigure}[t]{0.49\textwidth}
        \centering
        \includegraphics[width=\textwidth]{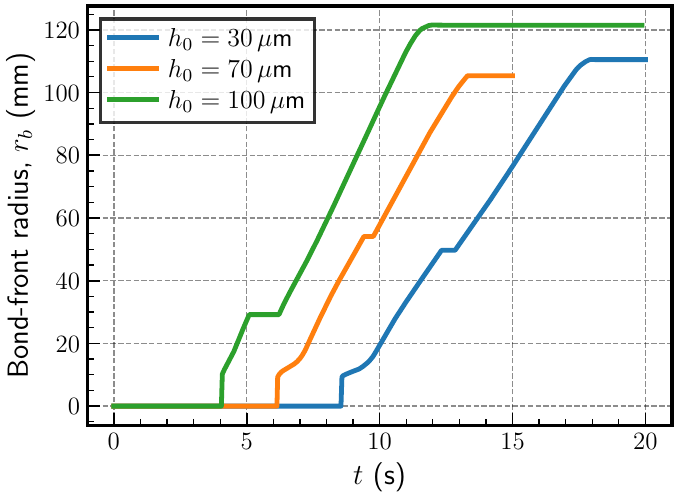}
        \caption{Bonded radius evolution}
        \label{fig:bondgap_radius_vs_time}
    \end{subfigure}
    \hfill
    \begin{subfigure}[t]{0.49\textwidth}
        \centering
        \includegraphics[width=\textwidth]{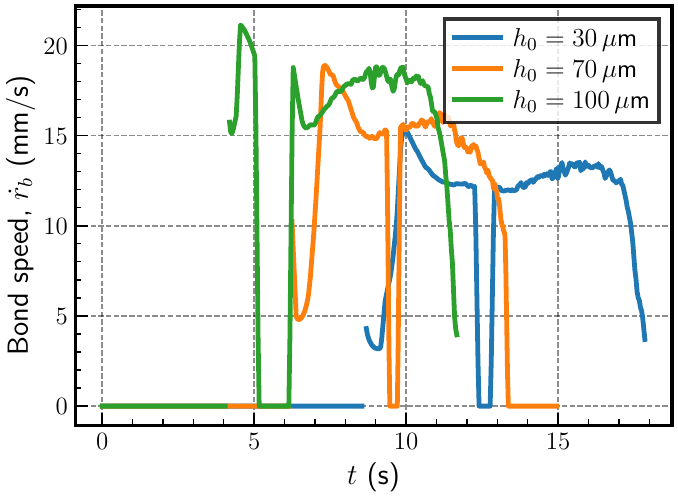}
        \caption{Bond speed versus time}
        \label{fig:bondgap_speed_vs_time}
    \end{subfigure}
    \par\vspace{0.6em}
    \begin{subfigure}[t]{0.49\textwidth}
        \centering
        \includegraphics[width=\textwidth]{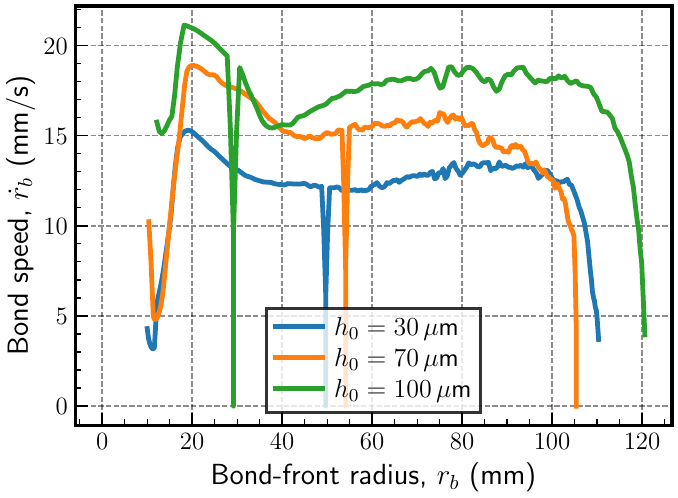}
        \caption{Bond speed along the radial coordinate}
        \label{fig:bondgap_speed_vs_radius}
    \end{subfigure}

    \caption{Effect of the initial bond gap for $h_0=30$, $70$, and $100\,\mu\mathrm{m}$: (a) bonded radius versus time, (b) bond speed versus time, and (c) bond speed versus radial position. Before bond initiation, $r_b=0$ and $\dot r_b=0$.}
    \label{fig:bondgap_speed}
\end{figure}

In this section, we examine the bonding speed for the three bond gaps considered in the validation exercise in Section~\ref{subsec:results_validation}. As shown in Fig.~\ref{fig:bondgap_speed}, the response is nonlinear and non-monotonic. Unlike bonding in the absence of air (pure plate bending), the largest bond gap reaches complete bonding earlier than the smallest gap; see Fig.~\ref{fig:bondgap_speed}(a). This ordering is driven by the air-film pressure $p_{\texttt{air}}$ in the unbonded region: smaller gaps generate higher $p_{\texttt{air}}$, which increases the resistance to bond-front advance as in demonstrated in Fig.~\ref{fig:air_pressure_profiles_tc}. The instantaneous bond-front speed remains nonlinear and is not ordered monotonically for all times, as illustrated in Figs.~\ref{fig:bondgap_speed}(b) and \ref{fig:bondgap_speed}(c). This behavior is expected, since the bond-front speed and the evolution of air pressure are governed by Reynolds' equation. Additional validation plots for the spatio-temporal displacement fields and air-pressure diagnostics are provided in Appendix~\ref{app:validation_more_results}; see Figs.~\ref{fig:heatmaps_displacement} and \ref{fig:air_pressure_profiles}.

\FloatBarrier

\subsection{Effect of Air Viscosity on air pressure and Bond Speed}
\begin{figure}[htbp]
    \centering
    \begin{subfigure}[t]{0.49\textwidth}
        \centering
        \includegraphics[width=\textwidth]{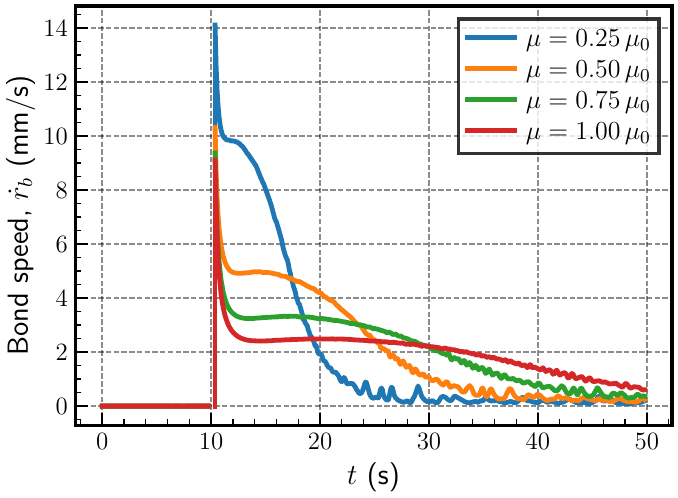}
        \caption{Bond speed versus time}
        \label{fig:viscosity_speed_vs_time}
    \end{subfigure}
    \hfill
    \begin{subfigure}[t]{0.49\textwidth}
        \centering
        \includegraphics[width=\textwidth]{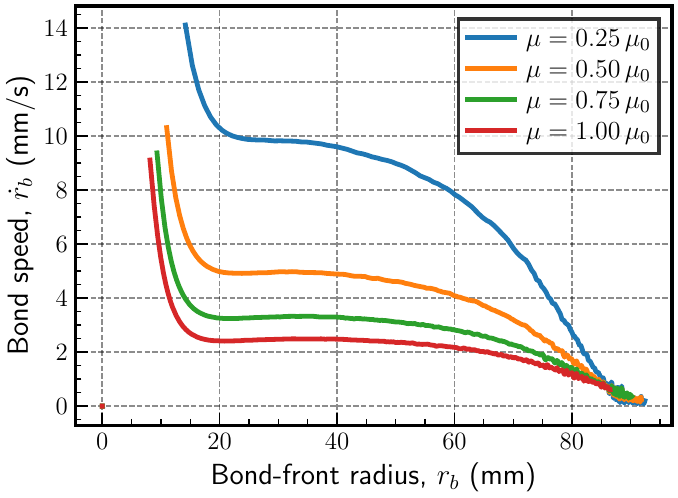}
        \caption{Bond speed along the radial coordinate}
        \label{fig:viscosity_speed_vs_radius}
    \end{subfigure}
    \par\vspace{0.6em}
    \begin{subfigure}[t]{0.49\textwidth}
        \centering
        \includegraphics[width=\textwidth]{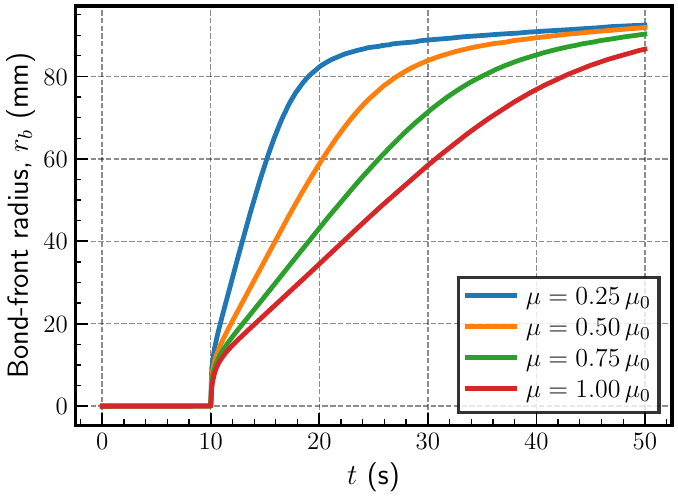}
        \caption{Bonded radius evolution}
        \label{fig:viscosity_radius_vs_time}
    \end{subfigure}

    \caption{Effect of air viscosity for $h_0=70\,\mu\mathrm{m}$ and $\mu/\mu_0=0.25$, $0.50$, $0.75$, and $1.00$: (a) bond speed over time, (b) bond speed versus radius, and (c) bonded-radius evolution, where $\mu_0 = 1.86\times10^{-5}\,\mathrm{Pa\cdot s}$ is the reference air viscosity.}
    \label{fig:viscosity_speed}
\end{figure}

Varying the air viscosity modifies the Reynolds pressure field and, consequently, the bond-front propagation rate. Figure~\ref{fig:viscosity_speed} summarises the post-processed diagnostics used to quantify this effect: bond speed versus time, the spatial variation of bond speed, and the temporal evolution of the bonded radius. For the same time instant, lower viscosity yields a faster bond front and a larger bonded area.

\FloatBarrier

\subsection{Effect of Interfacial Energy on bond speed}
\begin{figure}[htbp]
    \centering
    \begin{subfigure}[t]{0.49\textwidth}
        \centering
        \includegraphics[width=\textwidth]{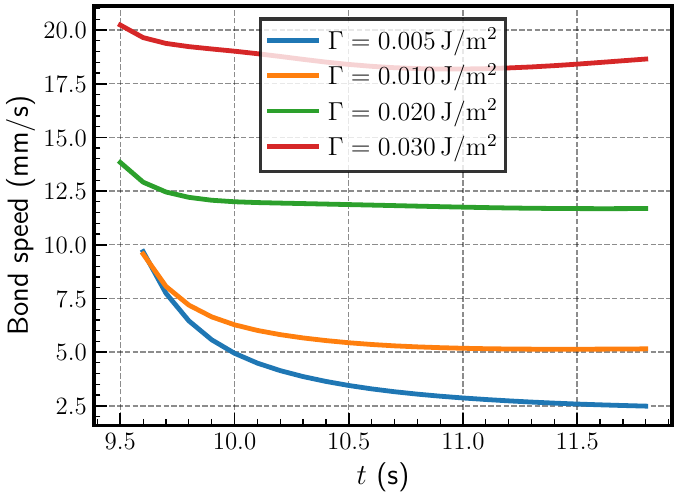}
        \caption{Bond speed versus time}
        \label{fig:interfacial_energy_speed_vs_time}
    \end{subfigure}
    \hfill
    \begin{subfigure}[t]{0.49\textwidth}
        \centering
        \includegraphics[width=\textwidth]{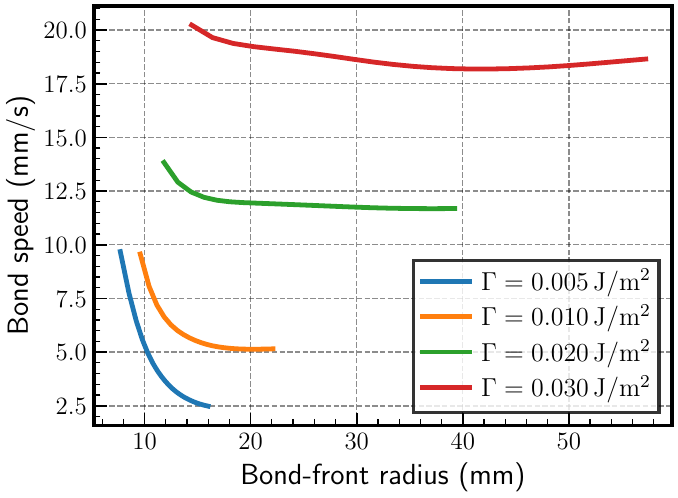}
        \caption{Bond speed along the radial coordinate}
        \label{fig:interfacial_energy_speed_vs_radius}
    \end{subfigure}
    \par\vspace{0.6em}
    \begin{subfigure}[t]{0.49\textwidth}
        \centering
        \includegraphics[width=\textwidth]{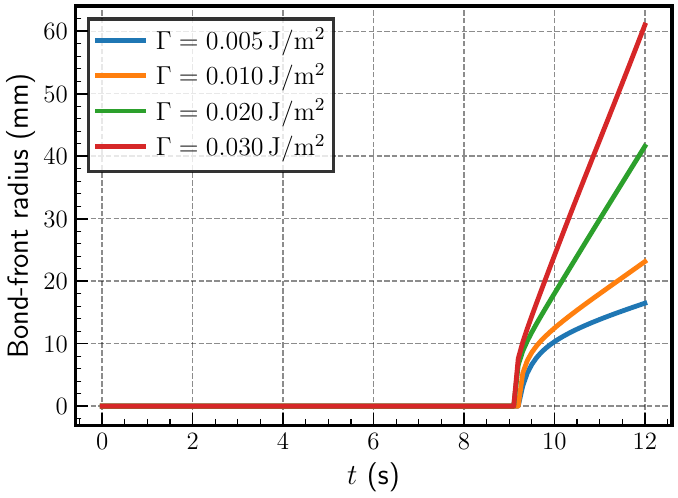}
        \caption{Bonded radius evolution}
        \label{fig:interfacial_energy_radius_vs_time}
    \end{subfigure}

    \caption{Effect of interfacial energy for $h_0=70\,\mu\mathrm{m}$ and $f_{\texttt{BE}}=0.005$, $0.010$, $0.020$, and $0.030\,\mathrm{MPa}$, corresponding to $\Gamma=0.005$, $0.010$, $0.020$, and $0.030\,\mathrm{J\,m^{-2}}$: (a) bond speed over time, (b) bond speed versus radius, and (c) bonded-radius evolution.}
    \label{fig:interfacial_energy_speed}
\end{figure}

Varying the interfacial energy $\Gamma$ changes the thermodynamic driving force for bonding, which in turn affects the bond speed. Figure~\ref{fig:interfacial_energy_speed} presents the diagnostics that capture this sensitivity: the bond speed as a function of time, the spatial distribution of the bond speed, and the temporal evolution of the bonded radius.


\section{Conclusion and Future Work}\label{sec:future}

\subsection{Conclusions}
This work developed a mathematically consistent reduced-order model for wafer-to-wafer bonding by coupling a Kirchhoff--Love plate description of wafer bending to a Reynolds lubrication model for the entrapped air film. Starting from three-dimensional linear elasticity, the plate equation was obtained in the thin-wafer limit and combined with a pressure evolution law in which the wafer deflection directly modulates the local gap height and flow resistance. The resulting nonlinear, strongly coupled system was discretized in a monolithic manner and implemented in the FEniCSx framework using a $C^0$ interior-penalty formulation for the fourth-order plate operator together with standard continuous Galerkin elements for the pressure field. The numerical studies demonstrated consistent force balance in the contact region and produced bonding-front dynamics in agreement with experimentally measured wafer - displacement histories, thereby providing a predictive and computationally efficient tool for assessing process sensitivities.

A key advantage of the present formulation is that the coupling structure is explicit: the air pressure enters the plate equation as a distributed transverse load, while the plate deflection enters the Reynolds equation through the evolving gap $h=h_0+w$. This structure enables systematic extensions of the physics without altering the overall solution strategy.

\subsection{Extension to anisotropic flexural response}
In the current isotropic plate model, the bending stiffness enters through the scalar flexural rigidity $D = Et_w^3/[12(1-\nu^2)]$. For crystalline wafers and layered stacks, however, the effective bending response can be direction-dependent. Within the same Kirchhoff--Love framework, anisotropy can be incorporated by replacing the scalar stiffness with a bending-stiffness tensor. In practice, one introduces the bending moment--curvature relation
\begin{align}
\bm{M} = \mathbb{D}\,\bm{\kappa},
\end{align}
where $\bm{M}$ and $\bm{\kappa}=\nabla^2 w$ are the bending-moment and curvature tensors, respectively, and $\mathbb{D}$ is a symmetric fourth-order bending stiffness (often expressed, under plane-stress reduction, as a $3\times 3$ matrix in Voigt notation). The plate equilibrium then becomes
\begin{align}
\nabla\cdot\nabla\!:\!\big(\mathbb{D}\,\bm{\kappa}\big) - q_{\texttt{ext}} + p_{\texttt{air}} - f_{\texttt{B}}(w) - f_c(w) = 0,
\end{align}
which reduces to $D\,\Delta^2 w$ when $\mathbb{D}$ corresponds to isotropic bending. Because the Reynolds equation and the coupling through $h$ remain unchanged, the anisotropic extension primarily modifies the plate operator and the constitutive evaluation of bending moments, and can be integrated into the present monolithic Newton framework with minimal changes.

\subsection{Future work}
Several extensions are planned. First, we will incorporate anisotropic and heterogeneous bending stiffness to represent crystallographic effects, device layers, and patterned stacks, enabling more accurate prediction of bonding kinetics for modern wafer architectures. Second, we will extend the reduced two-dimensional plate--Reynolds formulation to a geometrically richer shell model. A shell description will allow simultaneous prediction of (i) bond-front propagation and pressure evolution in the interfacial air film and (ii) in-plane deformations and post-bond overlays, including wafer stretch, shear, and in-plane distortion. This, in turn, will enable prediction of post-bond wafer shapes and residual distortions (e.g., global bow/warp and in-plane misfit patterns) that are critical for overlay, alignment, and downstream process yield. Finally, we will investigate the effect on bonding due to incoming wafer geometries.

\begin{appendices}
\section{Additional validation results}
\label{app:validation_more_results}

\begin{figure}[htbp]
    \centering
    \begin{subfigure}[t]{0.48\textwidth}
        \centering
        \includegraphics[width=\textwidth]{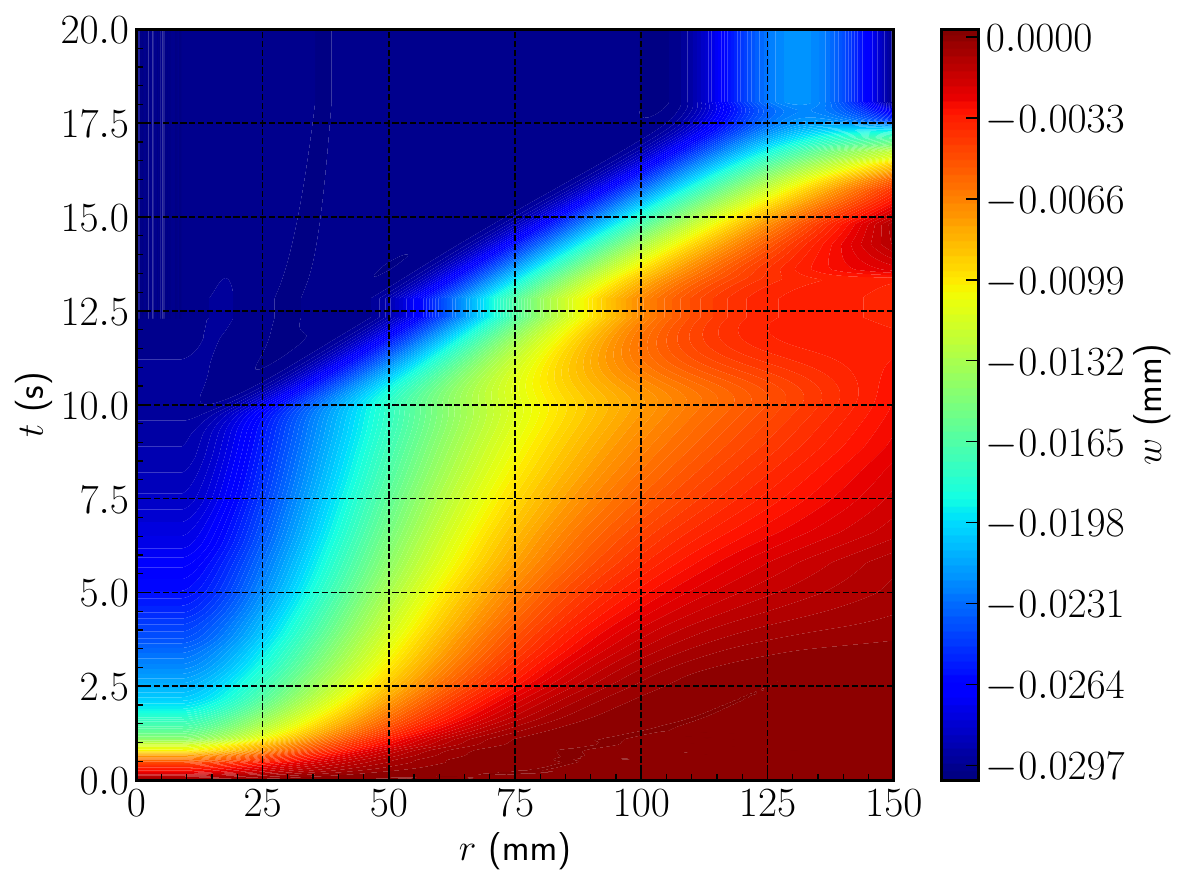}
        \caption{$h_0 = 30\,\mu$m}
        \label{fig:heatmap_disp_30}
    \end{subfigure}
    \hfill
    \begin{subfigure}[t]{0.48\textwidth}
        \centering
        \includegraphics[width=\textwidth]{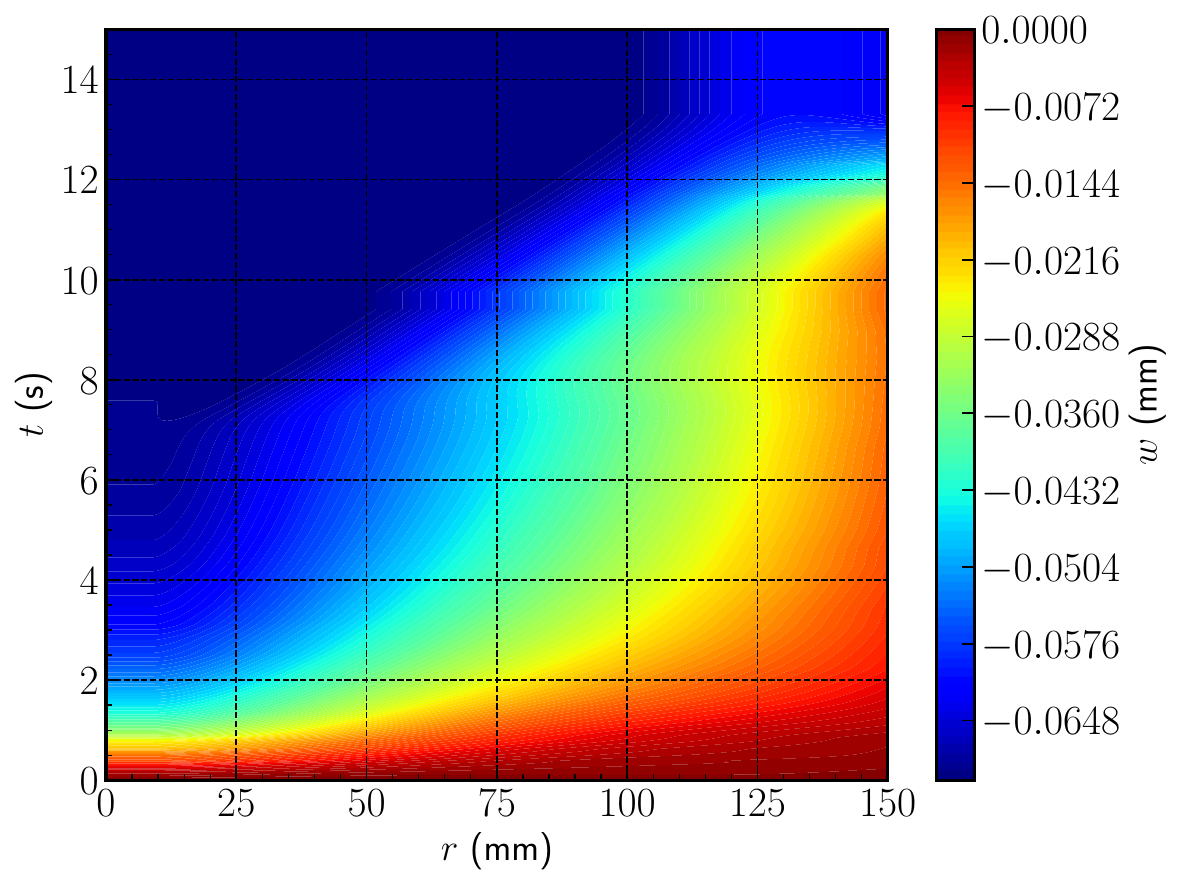}
        \caption{$h_0 = 70\,\mu$m}
        \label{fig:heatmap_disp_70}
    \end{subfigure}

    \vspace{0.5em}
    \begin{subfigure}[t]{0.48\textwidth}
        \centering
        \includegraphics[width=\textwidth]{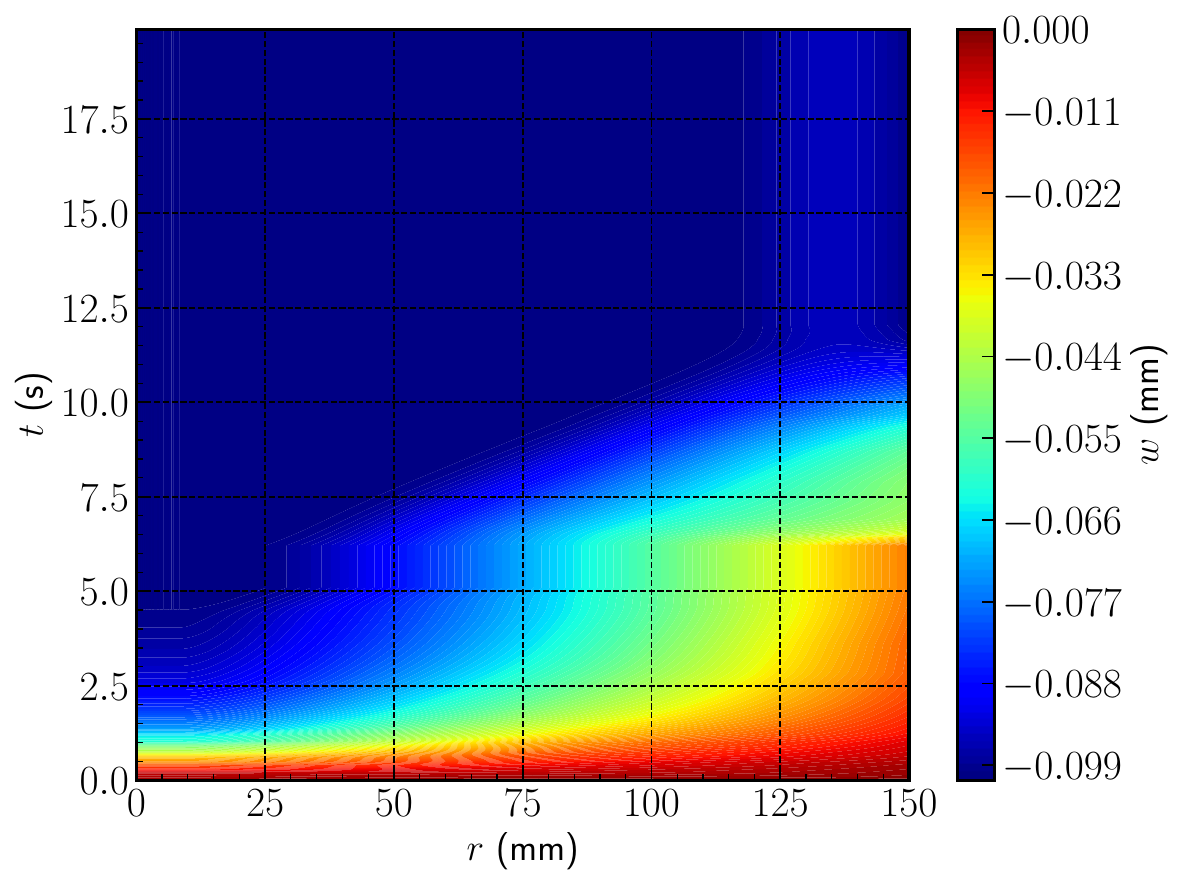}
        \caption{$h_0 = 100\,\mu$m}
        \label{fig:heatmap_disp_100}
    \end{subfigure}
    \caption{Spatio-temporal evolution of the radial displacement field $w(r,t)$ for three bond gaps.}
    \label{fig:heatmaps_displacement}
\end{figure}

\begin{figure}[htbp]
    \centering
    \begin{subfigure}[t]{0.48\textwidth}
        \centering
        \includegraphics[width=\textwidth]{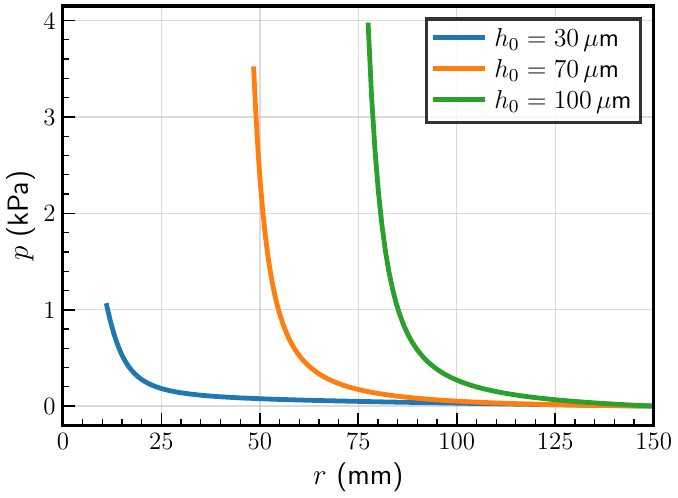}
        \caption{Full radial range}
        \label{fig:air_pressure_profiles_full}
    \end{subfigure}
    \hfill
    \begin{subfigure}[t]{0.48\textwidth}
        \centering
        \includegraphics[width=\textwidth]{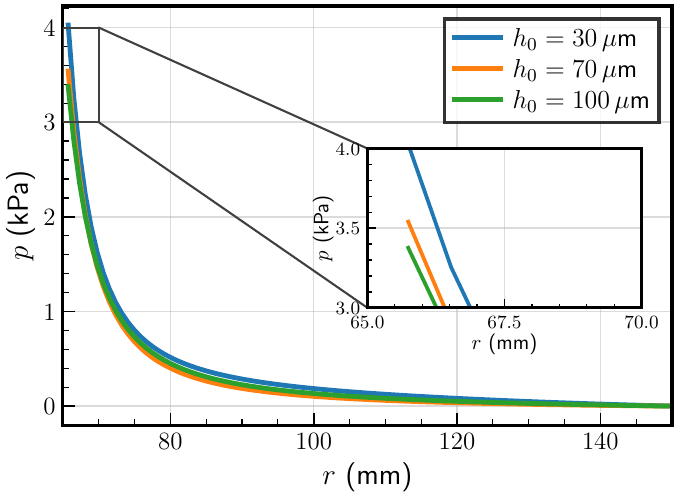}
        \caption{$p_{\texttt{air}}(r,t_c)$ at $r_b(t_c)=65\,\mathrm{mm}$}
        \label{fig:air_pressure_profiles_tc}
    \end{subfigure}
    \caption{Air-pressure diagnostics for $h_0=30$, $70$, and $100\,\mu\mathrm{m}$ for the result reported in Section~\ref{sec:bond_gap_bond_speed}: (a) radial profiles $p_{\texttt{air}}(r)$ in the unbonded region at $t=9\,\mathrm{s}$; and (b) profiles at time $t=t_c$ when $r_b(t_c)=65\,\mathrm{mm}$. The inset in (b) magnifies the bond-front region $65\leq r\leq70$.}
    \label{fig:air_pressure_profiles}
\end{figure}

This appendix \rev{section} collects additional plots supporting the validation study in Section~\ref{subsec:results_validation}. \rev{Figure~\ref{fig:heatmaps_displacement} presents the full spatio-temporal displacement field for each initial gap, from which the bond-front trajectories reported in Figure~\ref{fig:bondgap_speed}(a) are extracted. Figure~\ref{fig:air_pressure_profiles} presents the radial air-pressure profiles that explain the effect of bond gap on bond speed as discussed in Section~\ref{sec:bond_gap_bond_speed}.}

\section{\rev{Sensitivity to regularization parameter $\beta$}}
\label{app:beta_sensitivity}

\begin{figure}[htbp]
    \centering
    \includegraphics[width=0.9\textwidth]{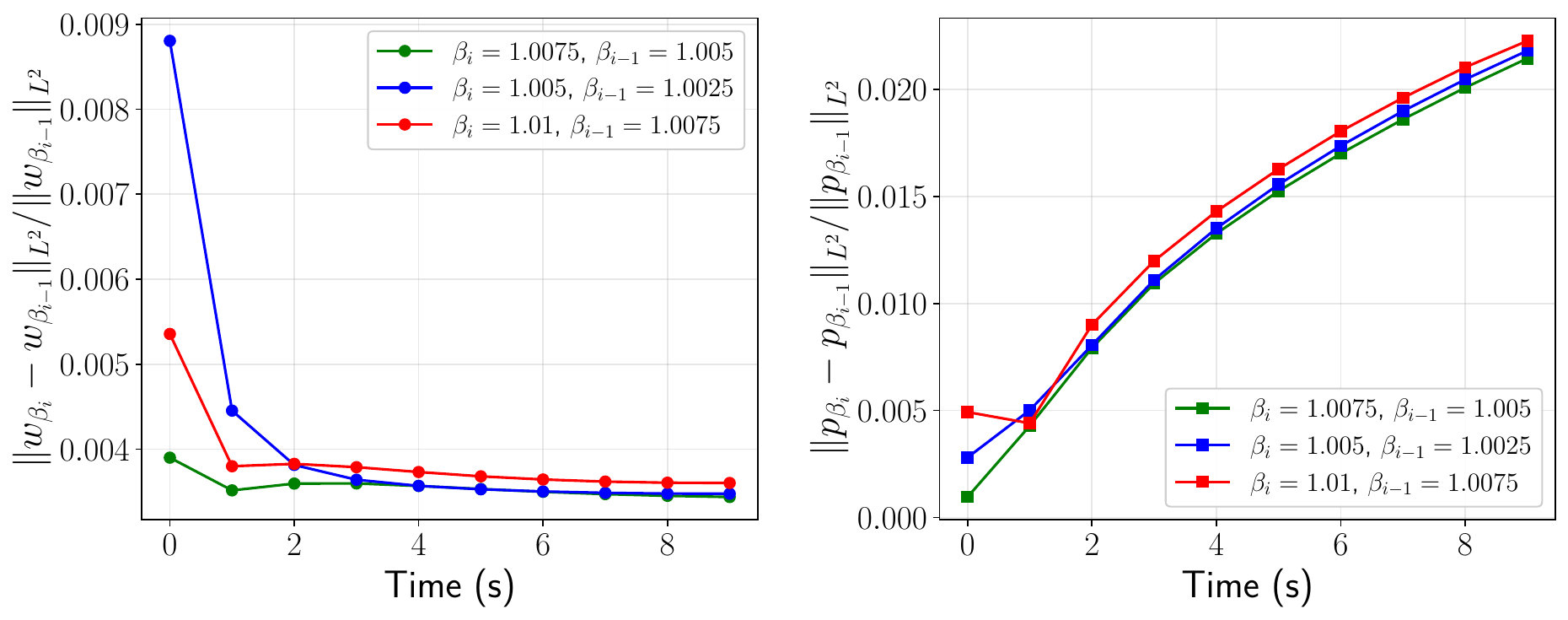}
    \caption{\rev{Effect of regularization parameter $\beta$ on successive difference norms for $h_0 = 70\,\mu\mathrm{m}$: (a) normalized deflection difference $\|w_{\beta_i} - w_{\beta_{i-1}}\|_{L^2}$ versus time, and (b) normalized pressure difference $\|p_{\beta_i} - p_{\beta_{i-1}}\|_{L^2}$ versus time, for consecutive pairs from $\beta \in \{0.0025, 0.005, 0.0075, 0.01\}$.}}
    \label{fig:beta_sensitivity}
\end{figure}

\rev{This appendix presents the sensitivity of the simulation results to the regularization parameter $\beta$ introduced in Section~\ref{subsec:strong_recap}. Simulations were performed for $h_0 = 70\,\mu\mathrm{m}$ with $\beta \in \{0.0025,\, 0.005,\, 0.0075,\, 0.01\}$. Figure~\ref{fig:beta_sensitivity} demonstrates that the displacement and pressure are approximately insensitive to $\beta$ within this range, confirming that the choice $\beta = 0.0075$ used throughout this work does not materially affect the reported results.}

\end{appendices}

\bibliography{references}
\bibliographystyle{unsrtnat}

\end{document}